\documentclass[AMA,LATO1COL]{WileyNJD-v2}
\usepackage{amsmath}
\usepackage{amssymb}
\usepackage{amsfonts}
\usepackage{color}
\usepackage{caption}
\graphicspath{{figures/}}
\usepackage{subfigure}
\usepackage[rel]{overpic}
\usepackage{comment}

\articletype{Article Type}

\received{<day> <Month>, <year>}
\revised{<day> <Month>, <year>}
\accepted{<day> <Month>, <year>}

\usepackage{titlesec}
\titlespacing*{\subsubsection} {3pt}{3pt}{3pt}
\titlespacing*{\subsection} {5pt}{5pt}{5pt}
\titlespacing*{\section} {12pt}{12pt}{12pt}

\usepackage{tikz,xcolor,hyperref}
\definecolor{lime}{HTML}{A6CE39}
\DeclareRobustCommand{\orcidicon}{%
	\begin{tikzpicture}
	\draw[lime, fill=lime] (0,0) 
	circle [radius=0.16] 
	node[white] {{\fontfamily{qag}\selectfont \tiny ID}};
	\draw[white, fill=white] (-0.0625,0.095) 
	circle [radius=0.007];
	\end{tikzpicture}
	\hspace{-2mm}
}
\foreach \x in {A, ..., Z}{%
	\expandafter\xdef\csname orcid\x\endcsname{\noexpand\href{https://orcid.org/\csname orcidauthor\x\endcsname}{\noexpand\orcidicon}}
}

\newcommand{\bsub}{\begin{subequations}}
\newcommand{\esub}{\end{subequations}}

\newcommand{\ep}{\epsilon}

\newcommand{\pt}{\mbox{\boldmath $\tau$}}

\newcommand{\bS}{\mathbf  S}

\newcommand{\pf}{\textbf{\emph{f}}}

\newcommand{\pu}{\textbf{\emph{u}}}

\newcommand{\pat}{\partial}
\newcommand{\na}{\nabla}
\newcommand{\x}{\times}

\newcommand{\beq}{\begin{equation}}
\newcommand{\eeq}{\end{equation}}
\newcommand{\bsubeq}{\begin{subequations}}
\newcommand{\esubeq}{\end{subequations}}
\newcommand{\beqn}{\begin{eqnarray}}
\newcommand{\eeqn}{\end{eqnarray}}
\newcommand{\fr}{\frac}
\newcommand{\lb}{\label}
\newcommand{\er}{\eqref}

\usepackage{hyperref}
\hypersetup{colorlinks,allcolors=blue}
\usepackage{amsmath}
\usepackage{multirow}
\usepackage{amsfonts,amssymb}
\usepackage{graphicx}
\usepackage{subfigure}
\usepackage[rel]{overpic}

\graphicspath{{figures/}}

\begin{document}

\title{Evaluating the accuracy of the actuator line model against blade element momentum theory in uniform inflow}

\author[1,2]{Luoqin Liu}
\author[3]{Lucas Franceschini}
\author[4]{Daniel F. Oliveira}
\author[5]{Flavio C. C. Galeazzo} 
\author[3]{Bruno S. Carmo}
\author[2]{Richard J. A. M. Stevens*}

\authormark{Liu \textsc{et al.}}

\address[1]{Department of Modern Mechanics, University of Science and Technology of China, Hefei 230026, Anhui, PR China}

\address[2]{Physics of Fluids Group, Max Planck Center Twente for Complex Fluid Dynamics, University of Twente, 7500 AE Enschede, The Netherlands}

\address[3]{Department of Mechanical Engineering, Escola Polit{\'e}cnica, University of S{\~a}o Paulo, S{\~a}o Paulo, SP 05508-030, Brazil}

\address[4]{Department of Naval Architecture \& Ocean Engineering, Escola Polit{\'e}cnica, University of S{\~a}o Paulo, S{\~a}o Paulo, SP 05508-030, Brazil}

\address[5]{High Performance Computing Center Stuttgart (HLRS), Nobelstr. 19, 70569 Stuttgart, Germany}

\corres{*Richard J. A. M. Stevens, Physics of Fluids Group, Max Planck Center Twente for Complex Fluid Dynamics, University of Twente, 7500 AE Enschede, The Netherlands. \\ \email{r.j.a.m.stevens@utwente.nl}}

\abstract[Abstract]{We evaluate the accuracy of the actuator line model (ALM) approach by performing simulations for the NREL~5~MW wind turbine in uniform inflow using three large eddy simulation codes. The power and thrust coefficients obtained using the three codes agrees within $1\%$ when the grid spacing $\Delta_{\rm grid} \le 5.25$~m, and are cross-validated against blade element momentum (BEM) theory. We find that the results of ALM converge towards BEM theory without the need for tip correction when the numerical resolution is increased. For $\Delta_{\rm grid}=0.98$~m the difference between the power and thrust coefficient obtained using ALM and BEM is $4.5\%$ and $2.1\%$, respectively, although we note that no absolute convergence between ALM and BEM can be obtained as both models use different assumptions, such as the use of a force projection method in the ALM. The difference in the local axial and tangential forces along the blades obtained from ALM simulations using $\Delta_{\rm grid} = 1.97$~m and $\Delta_{\rm grid} = 0.98$~m can be as large as $10\%$. The effect of the number of actuator points on the obtained turbine power and thrust coefficients is limited as the results converge when the spacing between the actuator points is about three times the grid spacing. This insight on the required number of blade points can be used to improve the efficiency of actuator line simulations.}


\maketitle

\section{Introduction}

Wind-turbine performance assessment and design are routinely performed using blade element momentum (BEM) theory \citep{gla26}. This method is a combination of the momentum theory introduced by Rankine \cite{ran65} from a macroscopic point of view and the blade element theory introduced by Froude \cite{fro78} from a local point of view. A basic assumption of the method is that half of the induction is over the rotor plane. Furthermore, the analysis assumes that the flow around the different blade sections is independent. The aerodynamic forces are obtained using tabulated aerofoil data derived from wind tunnel measurements or numerical simulations and corrected for three-dimensional effects \citep{sor02b}. However, owing to the limitations to represent various flow situations encountered in practice, it has become necessary to introduce different empirical corrections. Such situations include dynamic inflow, yaw misalignment, tip loss corrections, and heavily loaded rotors \citep{sor16}. Although BEM is sometimes referred to as a ``low-fidelity'' model, it should be noted that the BEM theory generates reliable results for rotors operating close to their design conditions and can therefore be used as a reliable reference when experimental data is not available \citep{dag20}.

Large eddy simulations (LES) have become a prominent tool for performing high-fidelity simulations of wind turbine wakes and wind farm flows \citep{tro10, ste17, mar18, dra18, por20}. LES can capture the three-dimensional unsteady character of the flow around wind turbines and evaluate the wind turbine performance. However, even with modern supercomputers, it is very challenging to fully resolve the flow around the blades in LES, and this requires the use of body-fitted meshes, see Refs.\ \citep{mit16,san17,gri21}. As the use of body-fitted meshes is not yet practical for wind farm simulations, the actuator line model (ALM) is widely used to model wind turbines. The ALM is based on the blade element theory using tabulated aerofoil data with the velocities computed from computational fluid dynamics (CFD) \citep{sor02b, iva07, mik07, tro10, tro11, por11, chu12, shi13, jha14, sor15}. 

The ALM was originally developed for horizontal axis wind turbines by S{\o}rensen and Shen \cite{sor02b} and has later been extended to vertical axis wind turbines \citep{abk17b}. In the ALM the forces on the blades, which are discretized as rotating lines of actuator points, are calculated using tabulated airfoil data based on the local angle of attack at the location of the actuator point at each time step of the calculation. Subsequently, the forces are projected from the actuator points to the flow solver to ensure the effect of the rotating blades is evaluated by the flow solver. Typically this force projection is performed using an isotropic Gaussian function \citep{sor02b}, although the use of non-isotropic Gaussian kernel function, in which the force projection width is different in each direction, has been proposed to represent the shape of the blades more accurately \citep{chu17, mar17, dra18, ma20}. Shives and Crawford \cite{shi13} suggested that the spreading parameter $\ep$ should be related to the local chord length $c$, namely as $c/8 \le \ep \le c/4$, for elliptical loaded wings. By minimizing the error between the potential flow around a two-dimensional airfoil and the flow generated by a Gaussian distributed body force Mart{\' i}nez-Tossas~et~al \cite{mar17} concluded that $\ep= 0.2c$ should be employed for a low angle of attack. Later, Rocchio~et~al \cite{roc20} found that at a high angle of attack potential flow theory is not appropriate as flow separation occurs, and the optimal value is $\ep \approx 0.1c$. Note that the highest resolution used in the present study is $\ep = 0.7 \bar c$ (case G in Table~\ref{tab.forceprojection}), where $\bar c=3.5$~m is the mean chord length of the blade. Therefore, in this work, we will not consider these aspects as even on modern supercomputers, the resolution of LES calculations is too limited to satisfy these conditions for wind farm simulations. Instead, we focus on a detailed analysis of the ALM accuracy in its {\it traditional} form to give more detailed analysis, insights, and guidelines of its accuracy for grid resolution employed in realistic, high-fidelity simulations of wind turbine wakes and wind farm flows.

This study considers the well-known National Renewable Energy Laboratory (NREL) 5~MW wind turbine, which was developed by Jonkman {\it et al.}\ \cite{jon09}, to provide a representative utility-scale reference turbine. This turbine has a rotor diameter of $D=126$~m, and the three blades are defined using the cross-sectional Delft University and National Advisory Committee for Aeronautics profiles since their aerodynamic properties are well known and are further extended to high angles of attack\citep{vit82} and corrected for three-dimensional rotational effects\citep{du98}, see also Ning\citep{nin13}. The obtained accuracy of the ALM model depends on the width of force projection compared to the grid spacing $\Delta_{\rm grid}$ and the spacing between the actuator points along the blade $\Delta_{\rm blade}$. It is well-known that the force projection $\ep$ needs to be sufficiently large to suppress numerical instabilities, while a large $\ep$ leads to inaccurate force distributions \citep{tro10, mar12, jha14, mar15b}. It is typically recommended to set $\ep \ge 2 \Delta_{\rm grid}$ \citep{tro10, mar15b}. Jha {\it et al.}\ \cite{jha14} proposed to use a grid resolution of $D/\Delta_{\rm grid} = [60, 120]$ in combination with $D/\Delta_{\rm blade} \ge 40$ to accurately capture the rotor thrust and power. In addition, they recommended that the actuator points should be chosen such that $\Delta_{\rm blade} \approx 1.5 \Delta_{\rm grid}$.
 
In this work, we reevaluate these guidelines by comparing three LES codes with BEM theory. In agreement with previous work, we find that $D/\Delta_{\rm grid}$ needs to be sufficiently high to get accurate results, in particular to get accurately capture the local forces along the blades. However, as a refinement of previous results, we will show that the number of actuator points is less important to capture the turbine's thrust and power accurately. It is shown that using $\Delta_{\rm blade} \leq 3 \Delta_{\rm grid}$ does not significantly improve the quality of the result. This indicates that the number of actuator points can be reduced compared to previous guidelines with {\it limited} loss of accuracy. This can be relevant when the ALM is employed in large-scale computations in which the ALM calculations, which are local in space, may result in significant computational overhead. The remainder of this paper is organized as follows. In section~\ref{sec.alm} we introduce the technical details of the ALM briefly. Section~\ref{sec.LES} introduces the numerical codes, and the main results and findings are presented in section~\ref{sec.results}. The conclusions are summarized in section~\ref{sec.conclusions}. 

\section{Actuator line model}\lb{sec.alm}
The turbine blades in the ALM are represented by distributed body forces on the quarter-chord lines of the blades. The velocity field is solved in a global coordinate system $(x, y, z)$ in streamwise, cross-stream, and vertical directions. The angle $\theta$ defines the azimuthal position of one of the blades, with the blades located at an angle $\Delta \theta = 2 \pi /B$ from each other, where $B$ is the number of the blades ($B=3$ for the NREL 5~MW turbine considered in this study). A local, rotating coordinate system $(r, \theta, x)$, as seen from the rotor blade, is used to determine the relative velocity. The lift and drag forces are calculated dynamically using the local velocity at each actuator point using tabulated drag and lift coefficients. 

Denoting $(u_x, u_y, u_z)$ the interpolated velocity on the actuator line points, $\Omega$ the rotor rotational speed, $r$ is the radius at the actuator line point, the local relative azimuthal velocity $u_\theta$ of the blade is then given as 
\beq 
u_\theta = \Omega r - u_y \cos \theta + u_z \sin \theta.
\eeq
The angle of attack $\alpha$ for each actuator point is given by
\beq
\alpha = \phi - \gamma, \quad 
\phi = \arctan \left( \fr{u_x}{u_\theta} \right),
\eeq
where $\gamma$ accounts for the local twist and pitch angle of the blade. The lift and drag forces per unit span are obtained using
\beq
f_L = \fr{1}{2} \rho u_{\rm rel}^2 c C_l, \quad 
f_D = \fr{1}{2} \rho u_{\rm rel}^2 c C_d, \quad 
\eeq
where $\rho$ is the density of fluid, $u_{\rm rel} = \sqrt{ u_\theta^2 + u_x^2 }$ is the local wind velocity relative to the blade, $c$ is the local chord length, and $C_l$ and $C_d$ are the local lift and drag coefficients corrected for three-dimensional effects, respectively. The forces are transferred from the rotor frame to the global coordinate frame as follows
\beqn 
f_x = - (f_L \cos \phi + f_D \sin \phi), \quad
f_y = - (f_L \sin \phi - f_D \cos \phi) \cos \theta, \quad
f_z = (f_L \sin \phi - f_D \cos \phi) \sin \theta.
\eeqn 

To avoid numerical instabilities caused by the turbine forces it is common to use a Gaussian force projection method \citep{sor02},
\beq\lb{eq.gauss}
\eta_\epsilon = \frac{1}{\epsilon^3 \pi^{3/2}} e ^{-d^2/\epsilon^2}, 
\eeq
with
\beq
d = \sqrt{ (x_{i,j,k}-x^{a})^2 + (y_{i,j,k}-y^{a})^2 +(z_{i,j,k}-z^{a})^2 }.
\eeq 
Here, the indices $i,j,k$ refer to the streamwise, spanwise, and vertical grid indices in the global coordinate frame, the superscript $a$ indicates the actuator line point index, 
and $\epsilon$ establishes the width of the force projection kernel. Based on the recommendation by Mart{\' i}nez-Tossas {\it et al.}\ \cite{mar15b}, we set $\epsilon = 2.5 \Delta_{\rm grid}$ throughout the entire study.
 
\begin{figure} 
\centering
\begin{overpic}[width=0.49\textwidth]{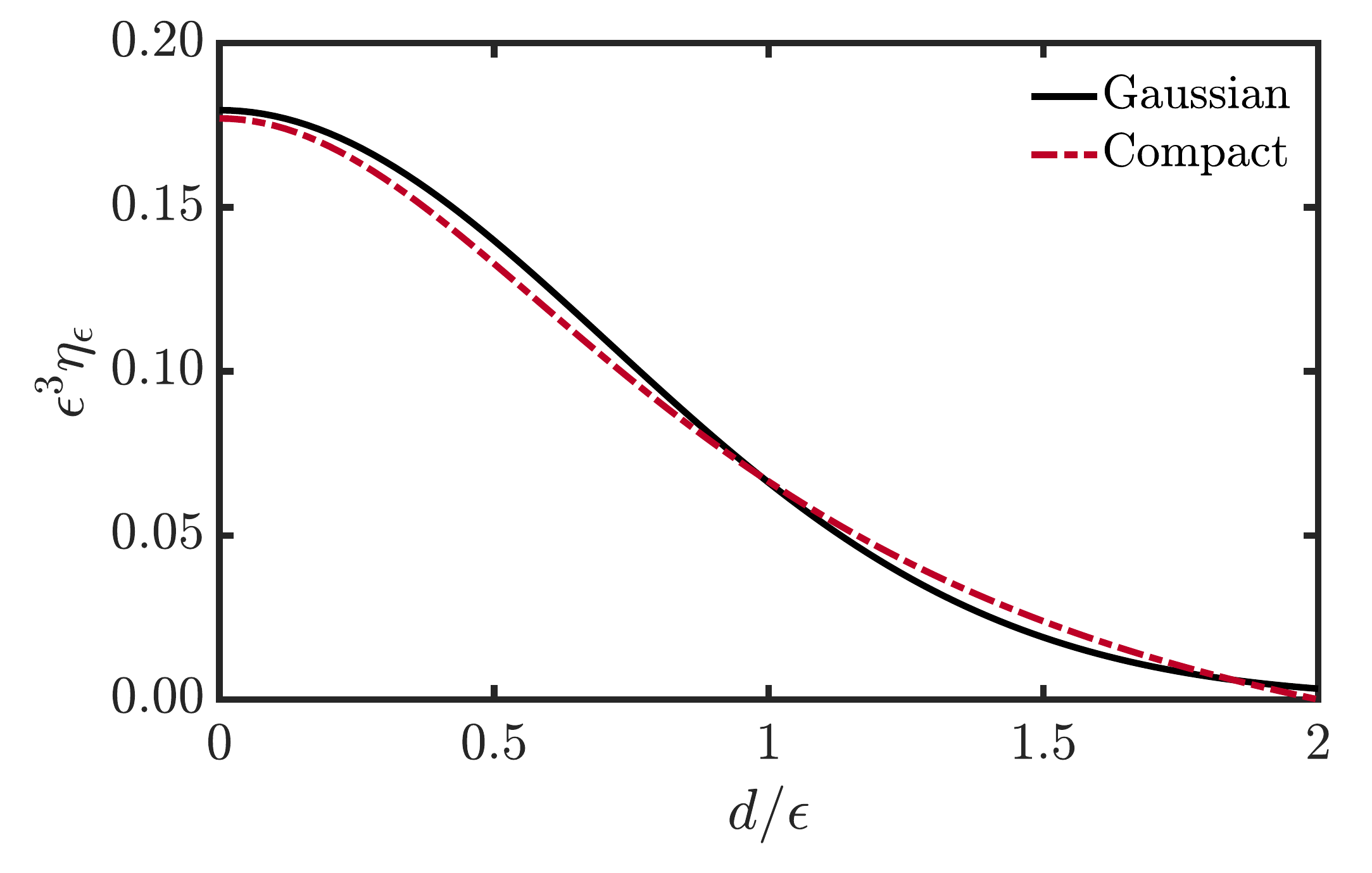} \end{overpic}
\vspace{-12pt} \caption{Comparison of the Gaussian kernel function \er{eq.gauss} and the compact kernel function \er{eq.compact} }
\label{fig.kernel}
\end{figure}

One downside of the Gaussian force projection method is that it goes to zero very slowly. This means that the region over which the force projection is calculated should be either truncated and the results normalized afterwards, or the calculation becomes very inefficient as the convolution is calculated over a very large region. A way to avoid this is to use the following compact force projection function 
\beq\lb{eq.compact}
\eta_\epsilon = \fr{a}{4 \pi \epsilon^3} \frac{ 4-(d/\ep)^2 }{ 1+(d/\ep)^2 } H(2-d/\ep), \quad 
a=\frac{3}{ 22-15 \arctan 2}, 
\eeq
where $H$ is the Heaviside function, 
\beq 
H(x)=0, \quad x<0; \quad 
H(x)=1, \quad x>0.
\eeq
It is shown in figure~\ref{fig.kernel} that this force projection function is very similar to the Gaussian projection function, which means that both force project methods should essentially give the same result. The results presented below confirm this. This force projection method is conservative over a distance of $2d/\epsilon$. This can provide computation benefits as the force projection radius that must be considered is confined and prevents the need to renormalize the results to account for small truncation errors required using the Gaussian projection method. 

\section{Brief overview codes}\lb{sec.LES}
	
In LES the spatially-filtered Navier-Stokes equations:
\beqn
\lb{eq.momentum}
\pat_t \widetilde{\pu} + \widetilde{\pu} \cdot \na \widetilde{\pu} = \pf_{\rm wt} - \na \widetilde {p} - \na \cdot \pt, \quad 
\na \cdot \widetilde{\pu} = 0,
\eeqn
where $\widetilde \pu$ is the velocity, $\pf_{\rm wt}$ is the force due to the wind turbine obtained using the ALM model described above, $\widetilde p$ is the modified pressure, and $\pt$ is the deviatoric part of the sub-grid scale shear stress, which is modeled as
\beq \lb {eq.t-ij}
\pt = - 2 \nu_t \widetilde {\bS}, \quad 
\nu_t = (C_{s} \Delta_{\rm grid})^2 \sqrt{ 2 \widetilde {\bS} : \widetilde {\bS} }, \quad 
\widetilde {\bS} \equiv \frac{1}{2} [ \na \widetilde {\pu} + (\na \widetilde {\pu})^T ].
\eeq
Here the superscript $T$ denotes a matrix transpose, $\Delta_{\rm grid} =(\Delta x \Delta y \Delta z)^{1/3}$ is the grid scale with $\Delta x, \Delta y, \Delta z$ the grid spacings in the streamwise ($x$), spanwise ($y$) and vertical ($z$) directions, respectively. All presented simulations are performed using the standard Smagorinsky model with $C_{s} = 0.16$ \citep{sma63}. Note that this value, as well as the turbulence model, has limited influence on the load distributions \cite{mar15b}.

The airfoil data\cite{jon09} is given at blade locations directed along the blade axis with coordinates at the midpoint of each blade element. The blade length and its root and tip locations are calculated using this information. Using the input parameters from Jonkman {\it et al.}\ \cite{jon09}, the total length of the wind turbine blades reported by WInc3D and turbinesFoam is smaller than the assumed 63 m due to the method that is used to calculate the blade element locations. The blade length is a relevant value, as a slightly shorter blade implies calculating the forces to be underpredicted. Both codes have been modified to account for this.

\subsection{Pseudo-spectral LES solver}

At the University of Twente we develop a pseudo-spectral LES code \citep{liu20, gad21} that is related to the LESGO code, which originates from the work by Albertson \cite{alb96}, and later contributions by Bou-Zeid {\it et al.}\ \cite{bou05}, Meyers and Meneveau \cite{mey10}, Calaf {\it et al.}\ \cite{cal10}, and Stevens {\it et al.}\ \cite{ste14}, Mart{\'i}nez-Tossas {\it et al.}\ \citep{mar15b, mar18}. The computational grids are uniformly distributed in the horizontal directions and are staggered in the vertical direction. The first vertical velocity grid plane is located at the ground, while the first horizontal velocity grid planes are located at half grid distance above the ground. The code is pseudo-spectral, implying periodic boundary conditions in the streamwise and spanwise directions. The vertical direction uses the second-order centered finite difference method. The boundary conditions in this direction are zero shear stress with no penetration. Time integration is done using the second-order Adams-Bashforth method.

\subsection{WInc3D}

WInc3D\citep{des20} provides an integrated wind farm simulation framework that allows detailed analyses of wake-wake and turbine-wake interactions. The code is based on higher-order compact finite-difference discretization schemes\citep{lai09} and uses an efficient 2D domain decomposition algorithm that allows the code to scale up to $O(10^5)$ computational cores\citep{lai11}. The pressure mesh is staggered from the velocity one by half a mesh to avoid spurious pressure oscillations. An explicit third-order Adams-Bashforth time advancement scheme is used for time marching. WInc3D offers several built-in models, including a native ALM that has been validated by comparisons with experiments\citep{des18,des19,des20}.

\subsection{turbinesFoam}

OpenFOAM \citep{wel98} is an open-source CFD toolbox, which coupled with the ALM library turbinesFoam \citep{bac18,bac19} allows the simulation of wind and marine hydrokinetic turbines. Functions such as interpolation, Gaussian projection, and vector rotation were adapted from NREL's SOWFA \cite{chu13-b}. The flow code is based on the finite volume method, with second-order accuracy in space and time. turbinesFoam has been validated for wind aligned and yawed tandem wind turbines\citep{one21}.

\section{Actuator line model results for NREL~5~MW in uniform inflow}\lb{sec.results}
We performed simulations for the NREL~5~MW reference wind turbine \citep{jon09} subject to non-turbulent uniform inflow condition of $U_\infty=8$~m/s, and no pressure gradient applied. The rotational speed of the rotor was fixed at $\Omega=9.1552$~RPM, giving a tip speed ratio of $\Lambda=7.55$. These conditions were chosen to provide the turbine's optimum power coefficient, extracting maximum energy from the flow. All simulations were performed in a $L_x \times L_y \times L_z = 8D \times 6D \times 6D$ domain in streamwise, spanwise and vertical directions, respectively. For simplicity, we used a uniform grid in all directions. The turbine was located at $3D$ downstream of the inlet in the middle of the $y$-$z$ plane. We used periodic boundary conditions in the spanwise direction and stress-free conditions in the vertical direction. The fringe region at the end of the domain was $7.5\%$ of the streamwise domain length \citep{ste14}. The results presented in this paper were obtained after the simulation reached its statistically stationary state. This test case was also used in previous validation studies; see e.g.\ Jha {\it et al.}\ \cite{jha14}, Mart{\' i}nez Tossas {\it et al.}\ \cite{mar15b}, Churchfield {\it et al.}\ \cite{chu17}, and Da{\v g} and S{\o}rensen \cite{dag20}.

\subsection{Blade element momentum analysis}\lb{sec.bem}

\begin{figure}[!t]
\centering
\begin{overpic}[width=0.49\textwidth]{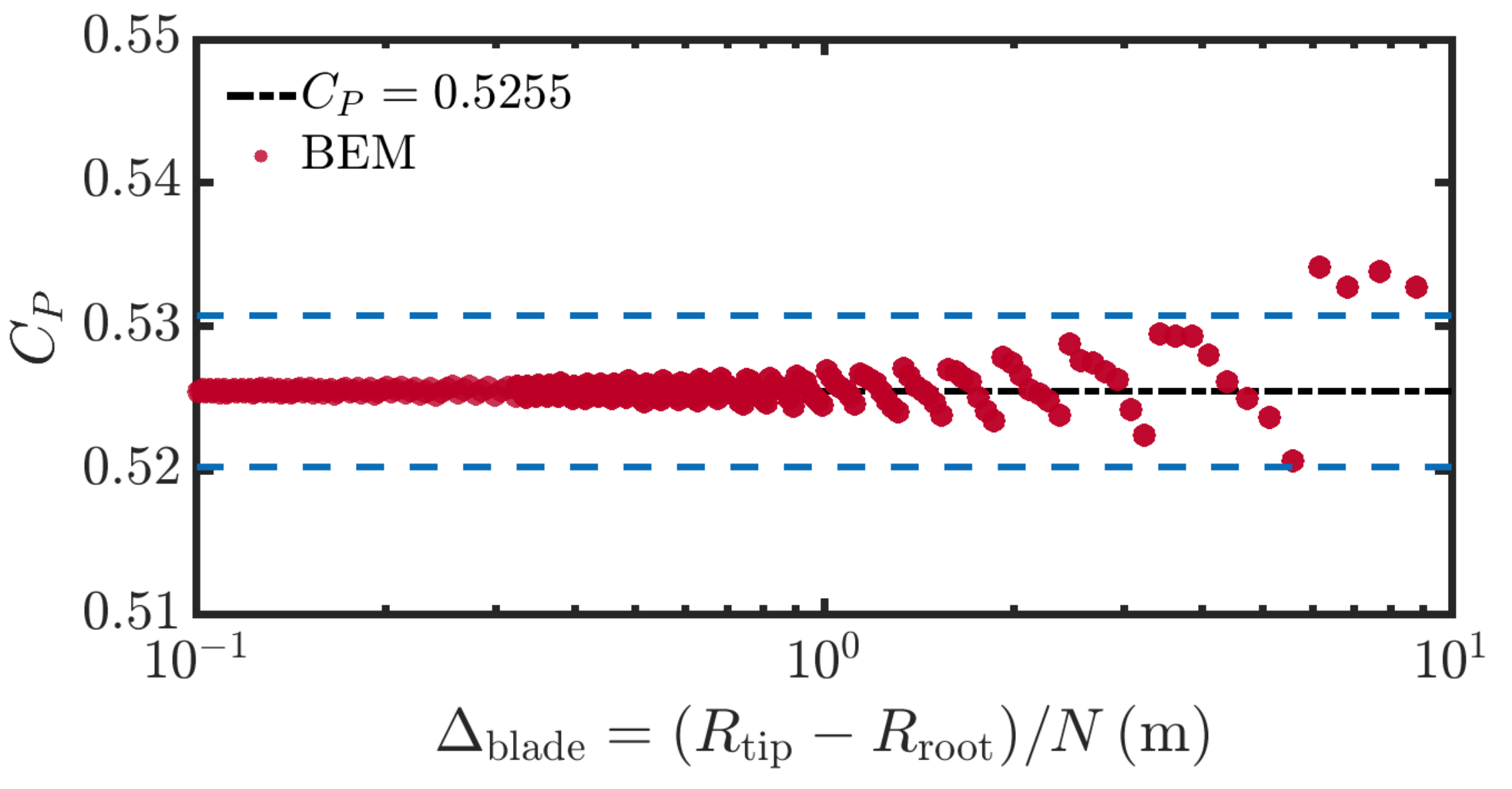}
\put(0,47){$(a)$}
\end{overpic}
\begin{overpic}[width=0.49\textwidth]{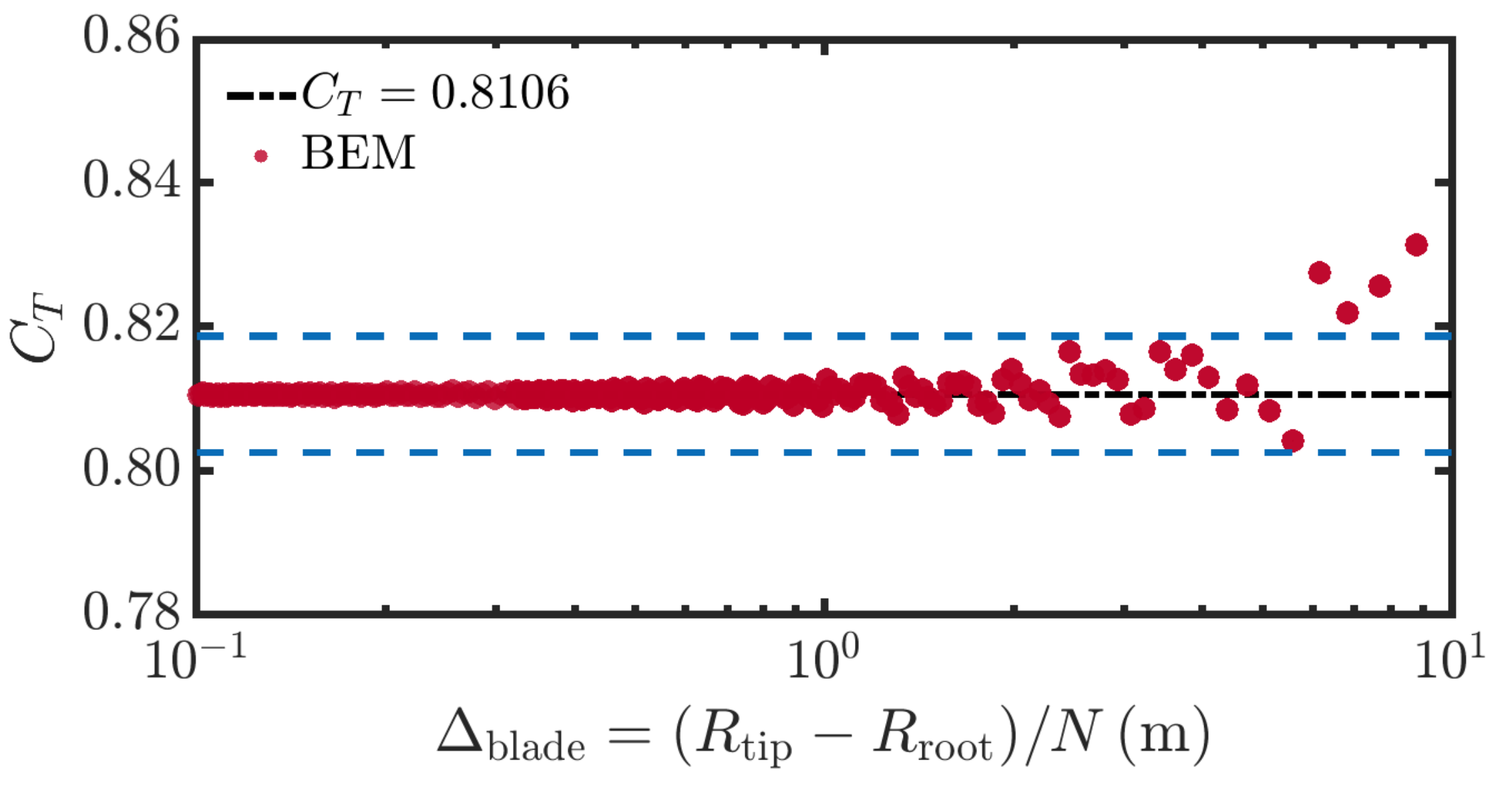}
\put(0,47){$(b)$}
\end{overpic}
\vspace{-12pt} \caption{The (a) power $C_P$ and (b) thrust $C_T$ coefficient obtained using BEM theory as function of the distance between the actuator points $\Delta_{\rm blade}$. The dashed-dotted black line indicates the high-resolution limit obtained using $N=1024$ actuator points. The dashed blue lines indicate the $1\%$ relative error, which is obtained for $N \ge 13$.}
\label{fig.bem-power}
\end{figure}

As the NREL~5~MW wind turbine is idealized such that there is no experimental data to compare with, we compare the LES results using the ALM with the theoretical predictions from BEM theory \citep{gla26, sor16}. For each blade element, the aerodynamic forces, similar to the ALM, are obtained using tabulated aerofoil data, which are assumed to be corrected for three-dimensional effects \citep{sor02, jon09}. We assume that the blade loading is not very heavy such that additional correction for the thrust coefficient is not required \citep{buh05}. Thus, the BEM analysis gives the following closed system \citep{bur01, sor16}, 
\beq
\tan \phi = \frac{1-a}{\lambda (1+a')}, \quad
\frac{ a }{ 1-a } = \frac{ \sigma C_n(\phi) }{4 \sin^2 \phi}, \quad
\frac{a'}{1-a} = \frac{\sigma C_t(\phi)}{4 \lambda \sin^2 \phi}.
\eeq
Here $\phi$ is the relative angle, $a$ and $a'$ are the axial and tangential induction factors, $\lambda = \Omega r / U_\infty$ is the local speed ratio, $\sigma = B c / (2\pi r)$ is the solidity of the turbine, and 
\beq
C_n(\phi) = C_l(\alpha) \cos \phi + C_d(\alpha) \sin \phi , \quad 
C_t(\phi) = C_l(\alpha) \sin \phi - C_d(\alpha) \cos \phi
\eeq
are the normal and tangential force coefficients, respectively, where $\alpha=\phi-\gamma$ is the local angle of attack and $\gamma$ is the blade twist angle, and $C_l$ and $C_d$ are the corrected lift and drag coefficients. The closed system $(\phi, a, a')$ can be solved by an iterative method \citep{bur01, sor16}, of which the existence and uniqueness have been proved mathematically by Ledoux {\it et al.}\ \cite{led20}. Subsequently, the axial and tangential forces per unit span along each blade can be obtained 
\beq
F_n = \fr{1}{2} \rho U_\infty^2 C_n c \left( \fr{ 1-a } { \sin \phi } \right)^2, \quad 
F_t = \fr{1}{2} \rho U_\infty^2 C_t c \left( \fr{ 1-a } { \sin \phi } \right)^2.
\eeq 
Finally, the total thrust $T$ and power output $P$ of the wind turbine can be obtained by integrating these forces along the three turbine blades as follows
\beq
T = \fr{3}{2} \rho U_\infty^2 \int C_n c \left( \fr{ 1-a } { \sin \phi } \right)^2 {\rm d} r, \quad 
P = \fr{3}{2} \rho U_\infty^2 \Omega \int C_t c \left( \fr{ 1-a } { \sin \phi } \right)^2 r {\rm d} r.
\eeq 
Figure~\ref{fig.bem-power} shows the power coefficient $C_P=8P/(\rho U_\infty^3 \pi D^2)$ and the thrust coefficient $C_T=8T/(\rho U_\infty^2 \pi D^2)$ obtained using BEM theory as function of the distance between the actuator points $\Delta_{\rm blade} = (R_{\rm tip}-R_{\rm root})/N$, where $N$ is the number of actuator points along each blade. The figure shows that the $C_P$ and $C_T$ values obtained using $\Delta_{\rm blade} \le 4.7$~m, which corresponds to the use of $N\ge 13$ actuator points along each blade, agree within 1\% with the high-resolution limit obtained by $N=1024$.

\begin{table}[!t]%
\centering
\caption{Comparison of power $C_P$ and thrust $C_T$ coefficients obtained in the UTwente-LES simulations with the Gaussian and compact force projection methods. The columns from left to right indicate the case name, the used numerical resolution in the streamwise, spanwise, and vertical direction ($N_x \times N_y \times N_z$), the grid spacing $\Delta_{\rm grid}$. The number of actuator points along each blade is $N=N_y/3$. The $C_{P, g}$ and $C_{T, g}$ are obtained using the Gaussian projection function (Eq.~\eqref{eq.gauss}) and $C_{P, c}$ and $C_{T, c}$ using the compact projection function (Eq.~\eqref{eq.compact}).}
\label{tab.forceprojection}
\begin{tabular*}{500pt}{@{\extracolsep\fill}ccccccc@{\extracolsep\fill}}
\toprule
Case & $N_x \times N_y \times N_z$ & $\Delta_{\rm grid}$ (m) & $C_{P,g}$ & $C_{P,c}$ & $C_{T,g}$ & $C_{T,c}$ \\
\midrule
A & $128 \times 96 \times 96$ & 7.88 & 0.6034 & 0.6002 & 0.8542 & 0.8521 \\
B & $192 \times 144 \times 144$ & 5.25 & 0.6077 & 0.6051 & 0.8618 & 0.8602 \\
C & $256 \times 192 \times 192$ & 3.94 & 0.5891 & 0.5868 & 0.8521 & 0.8505 \\
D & $384 \times 288 \times 288$ & 2.63 & 0.5745 & 0.5727 & 0.8444 & 0.8433 \\
E & $512 \times 384 \times 384$ & 1.97 & 0.5650 & 0.5633 & 0.8379 & 0.8368 \\
F & $768 \times 576 \times 576$ & 1.31 & 0.5547 & 0.5531 & 0.8313 & 0.8302 \\
G & $1024 \times 768 \times 768$ & 0.98 & 0.5494 & 0.5479 & 0.8280 & 0.8270\\ \bottomrule
\end{tabular*}
\end{table}

\subsection{Effect of the force projection function}\lb{sec.forceprojection}

Here we present a comparison of the results obtained using different methods to account for the presence of actuator points in the flow, namely, the Gaussian force projection method, Eq.~\er{eq.gauss}, and the compact force projection function, Eq.~\er{eq.compact}. From the comparison between the two force functions presented in figure~\ref{fig.kernel}, we did not expect significant differences in the $C_P$ and $C_T$ results obtained with either approach and table~\ref{tab.forceprojection} shows this is what indeed happens. In particular, table~\ref{tab.forceprojection} shows that the newly proposed force projection method gives essentially the same results (the <1\% difference is less than other uncertainties) as the commonly employed Gaussian force projection method. Hence all the findings obtained for the Gaussian force projection method extend to the newly introduced force projection.

The data in table~\ref{tab.forceprojection} were calculated with the pseudo-spectral LES solver, but we also ran tests using WInc3D, which also demonstrated that the differences are minimal, thus confirming our findings. So this shows that the computational performance of ALM codes can possibly be improved by using this alternative formulation, since the operations are cheaper than the traditional Gaussian projection approach, without any noticeable change in the results within a tolerance of less than 1\%. We remark that the computation benefit of the new force projection is expected to depend on the considered case, for example, on the number of turbines compared to the considered volume and the used simulation code. In the present test case, we consider only one turbine, and therefore the benefit of the proposed force projection method is limited (a few percentage points) as the relative computational requirements for the ALM calculations are limited for this case. In any case, we note that reducing the number of actuator blade points, as suggested by our findings below, is a more effective way to reduce the computational overhead of the ALM calculations. However, a benefit of the compact function is that no truncation is required, preventing small fluctuations in time-dependent forces as the blades move through the grid.

\subsection{Effect of the LES grid spacing}\lb{sec.grid}

\begin{figure}[!t]
\centering
\begin{overpic}[width=0.4\textwidth]{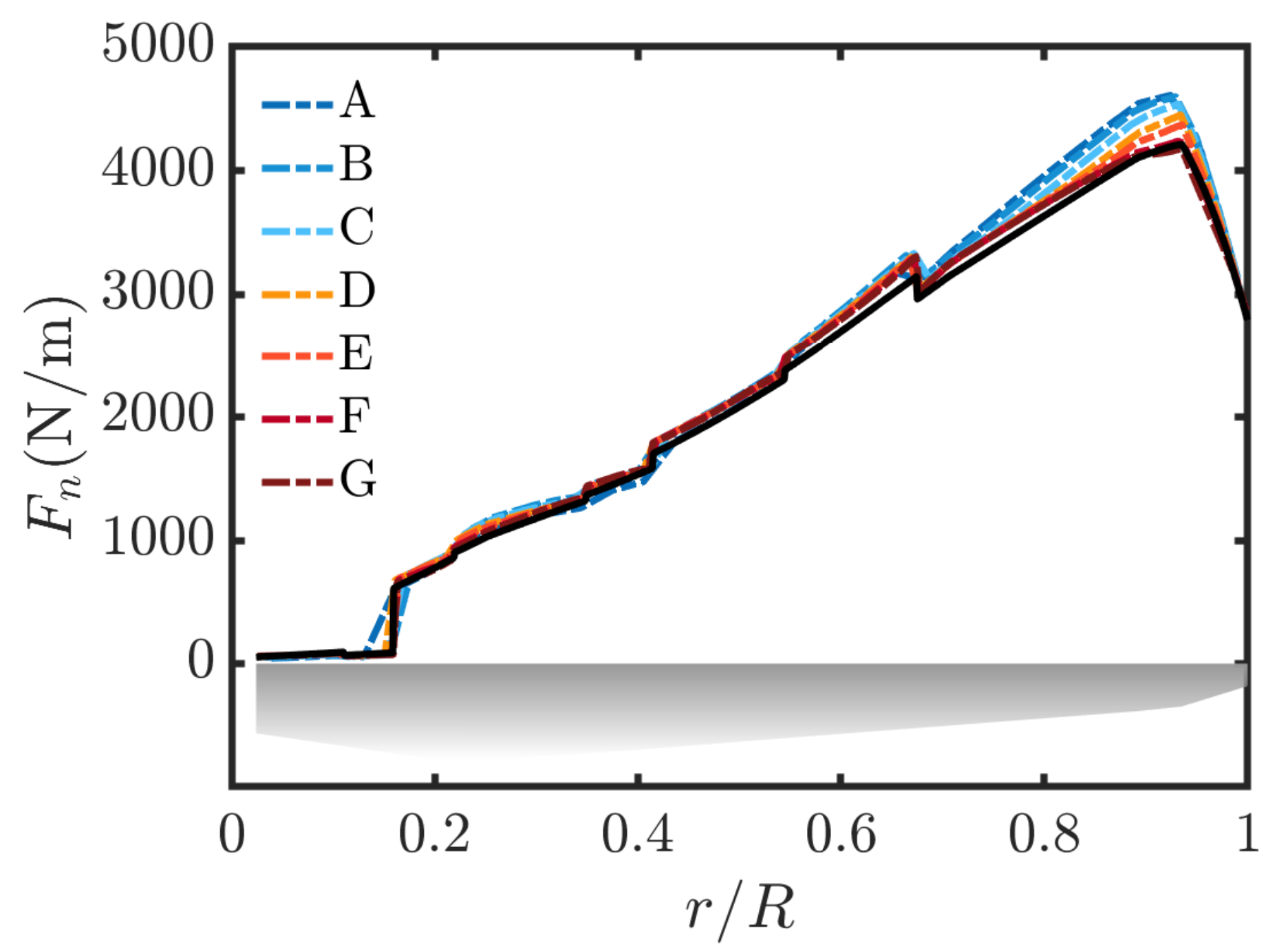}
\put(0,67){$(a)$}
\end{overpic} 
\begin{overpic}[width=0.4\textwidth]{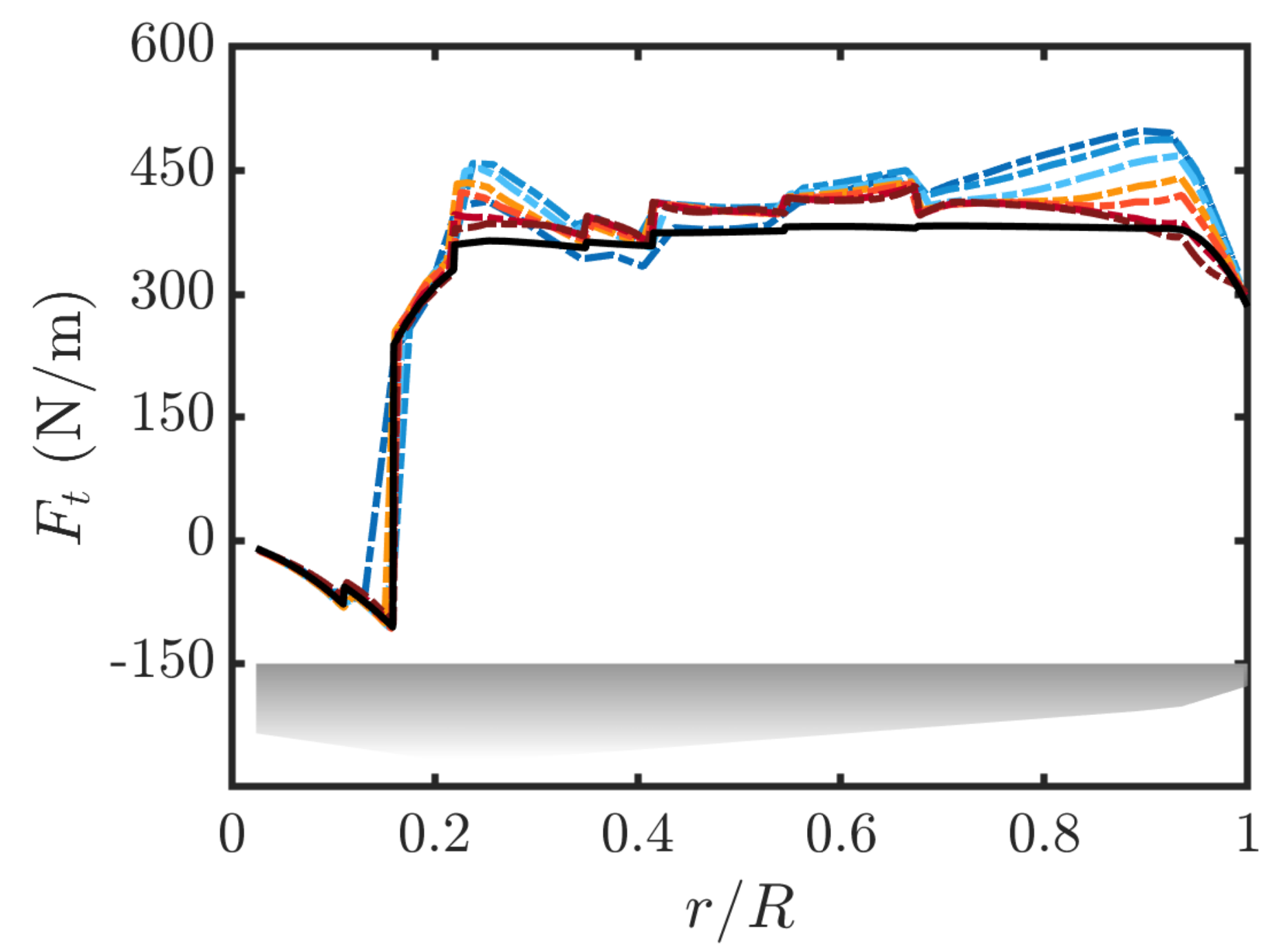}
\put(0,67){$(b)$}
\end{overpic}
\begin{overpic}[width=0.4\textwidth]{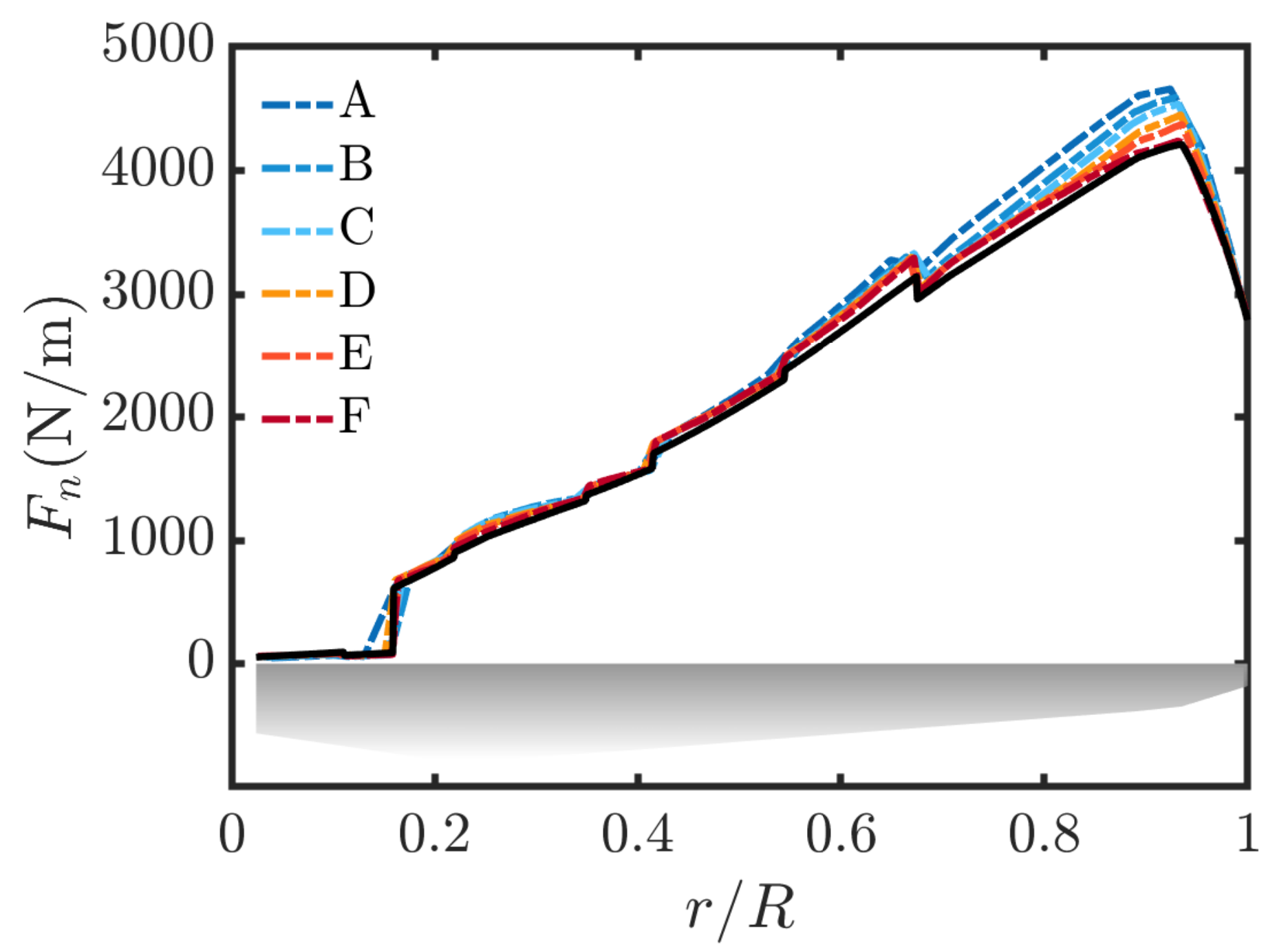}
\put(0,67){$(c)$}
\end{overpic} 
\begin{overpic}[width=0.4\textwidth]{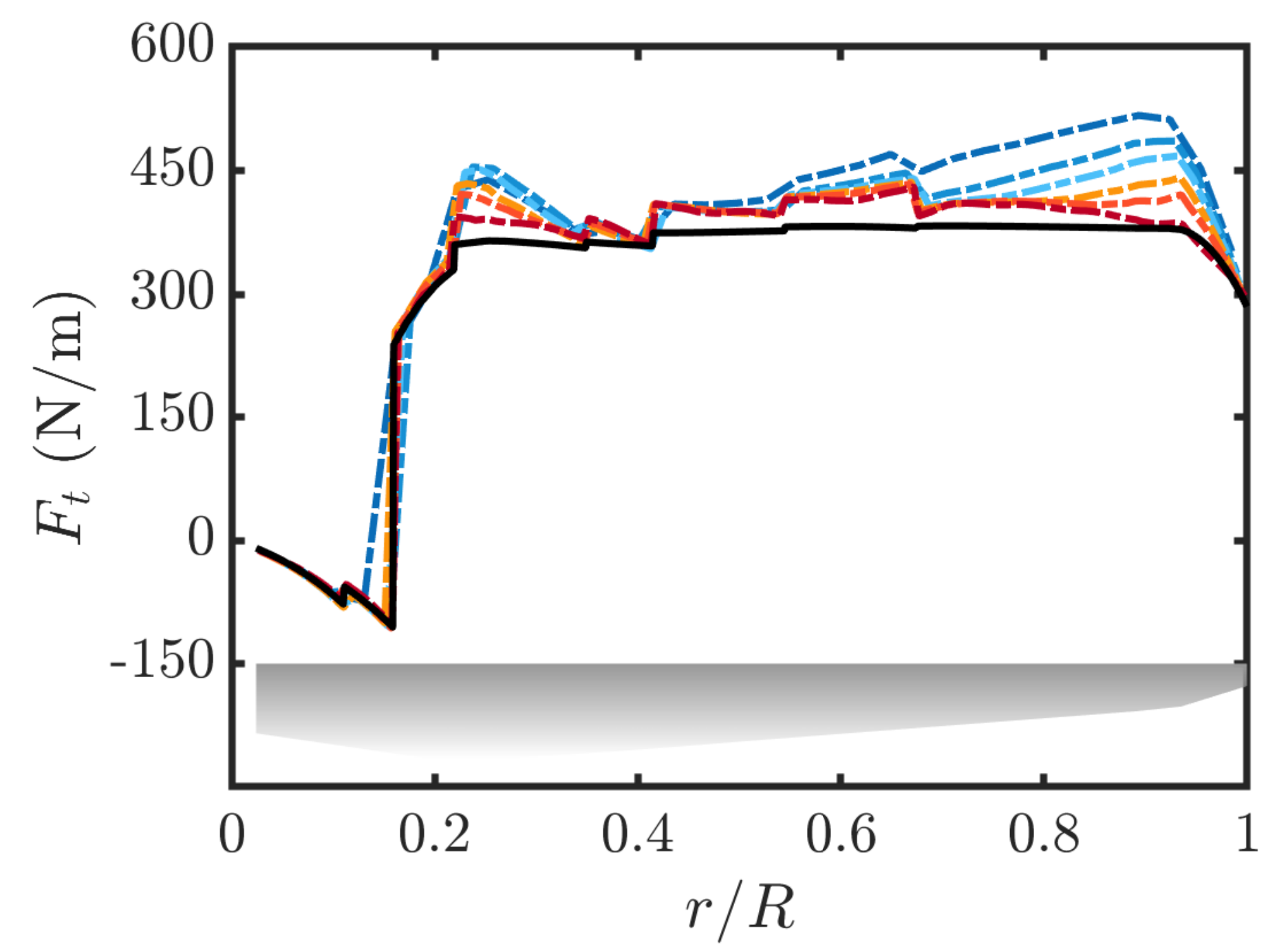}
\put(0,67){$(d)$}
\end{overpic}
\begin{overpic}[width=0.4\textwidth]{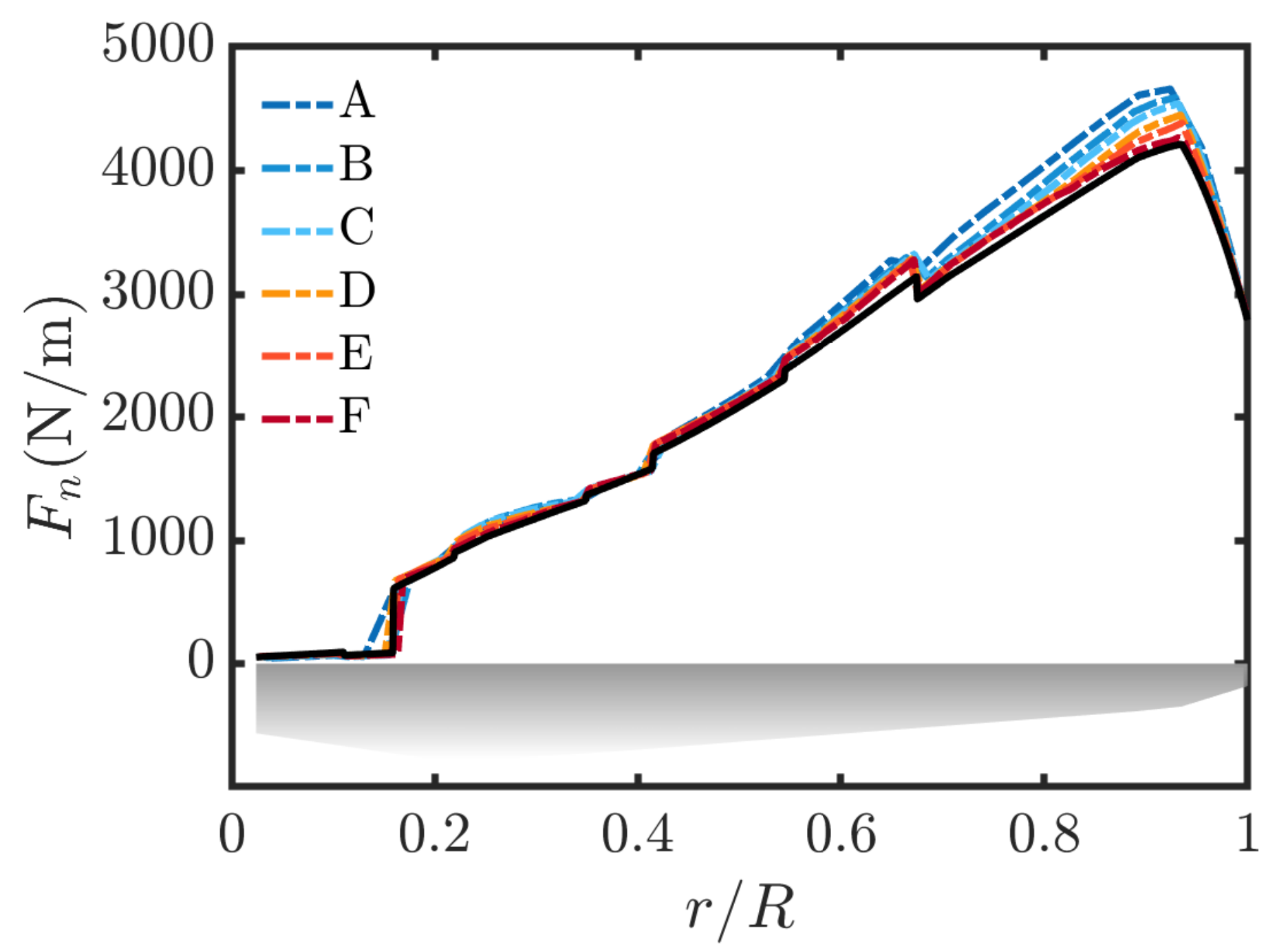}
\put(0,67){$(e)$}
\end{overpic} 
\begin{overpic}[width=0.4\textwidth]{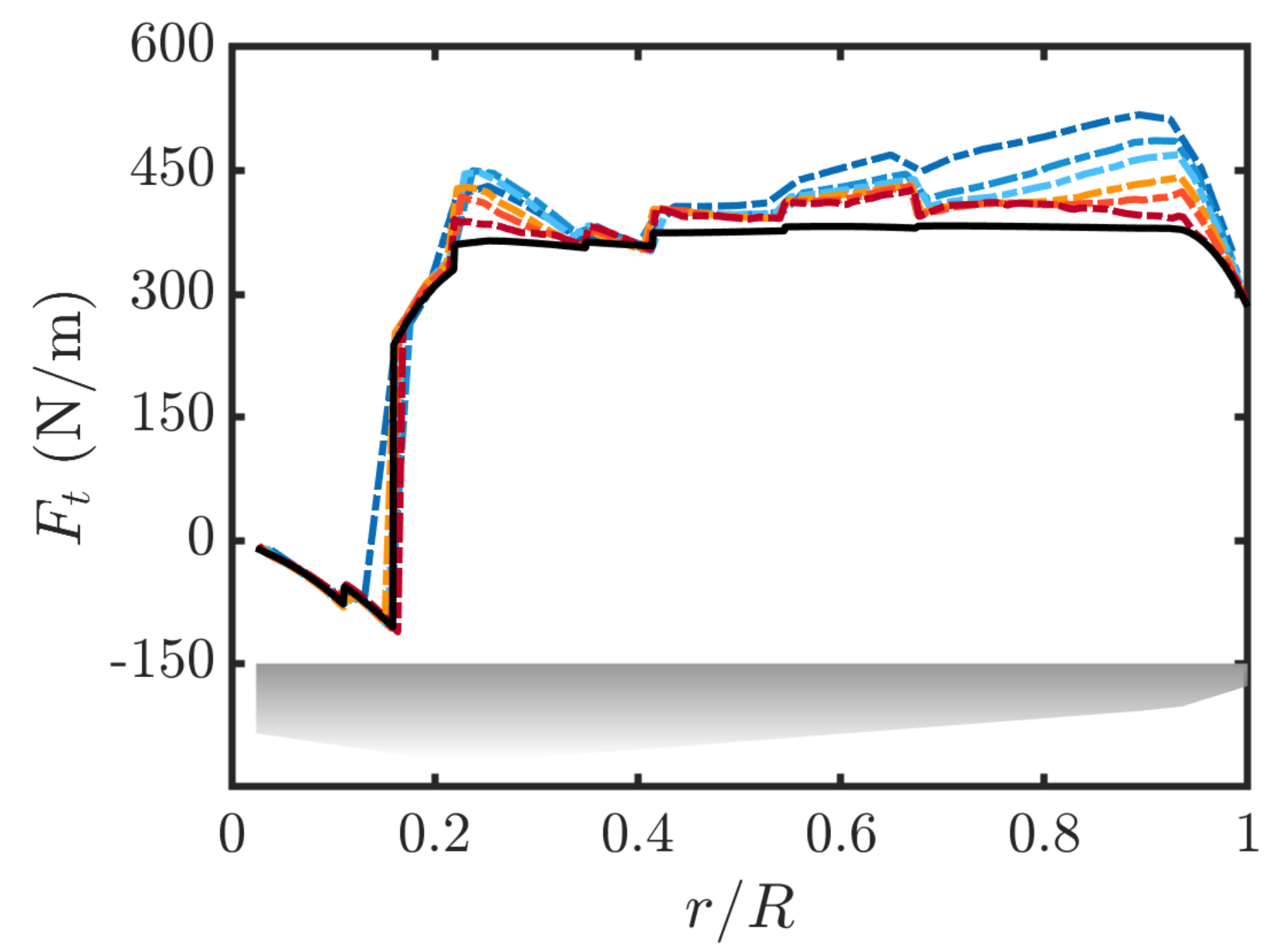}
\put(0,67){$(f)$}
\end{overpic}
\vspace{-12pt} \caption{The (a,c,e) axial and (b,d,f) tangential force per unit span along the blade obtained in the pseudo-spectral LES solver (a,b), Winc3D (c,d), and turbinesFoam (e,f) simulations. The solid line indicates the high-resolution ($N=1024$) BEM reference results and the symbols indicate the simulation results obtained on different grid resolutions, see table \ref{tab.grid} for details.}
\label{fig.grid-blade1}
\end{figure}

\begin{figure}[!t]
\centering
\begin{overpic}[width=0.4\textwidth]{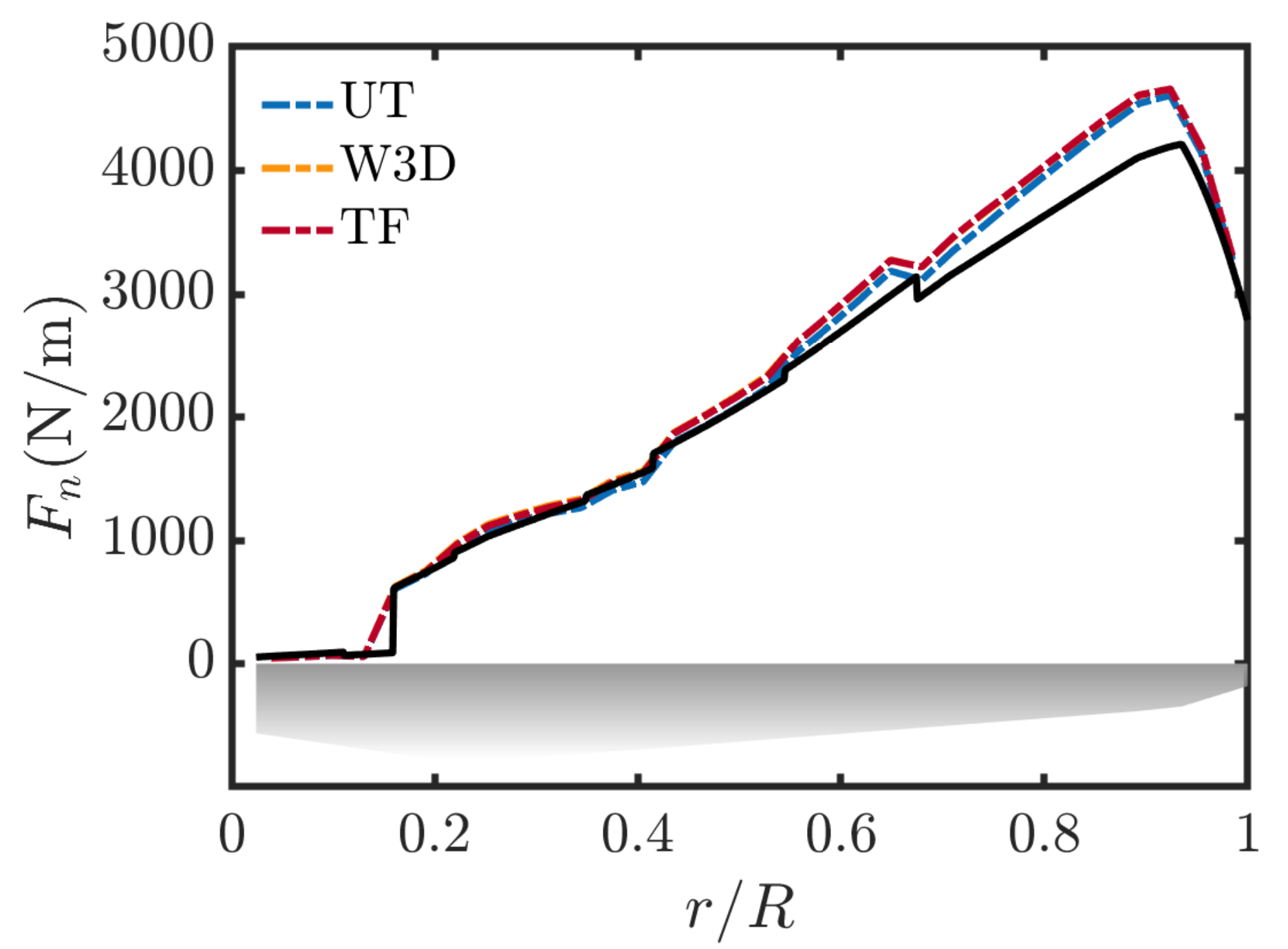}
\put(0,67){$(a)$}
\end{overpic} 
\begin{overpic}[width=0.4\textwidth]{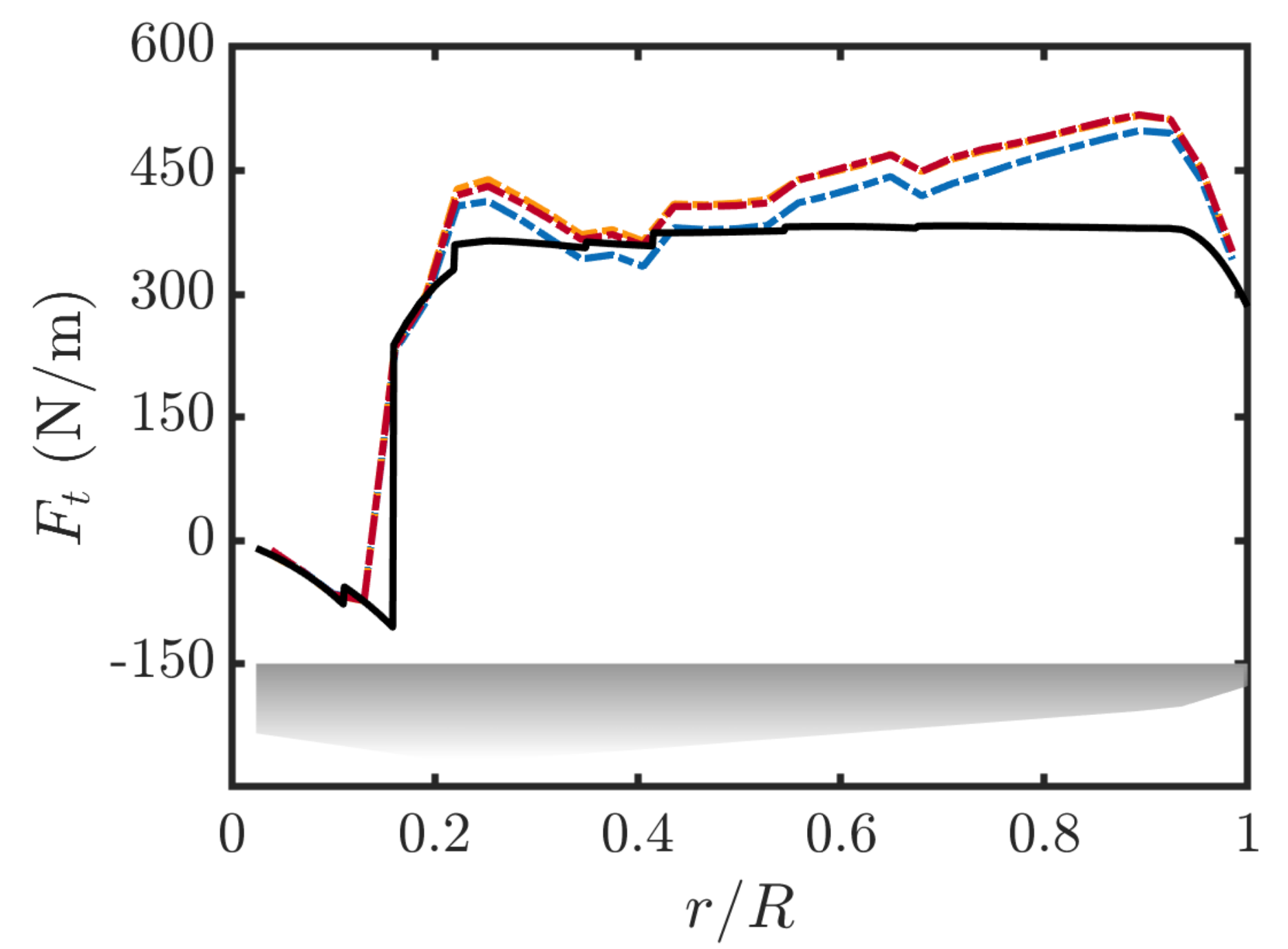}
\put(0,67){$(b)$}
\end{overpic}
\begin{overpic}[width=0.4\textwidth]{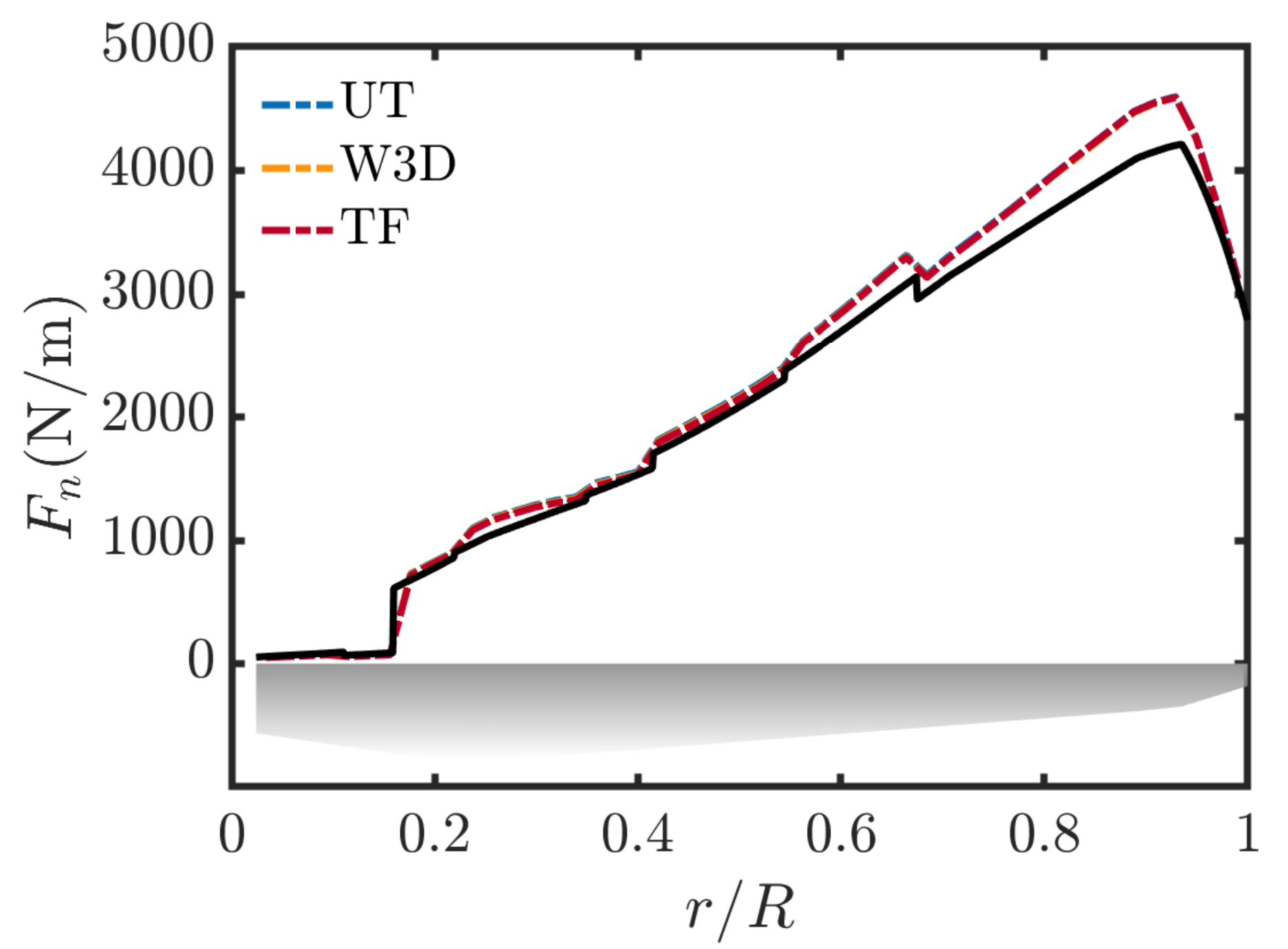}
\put(0,67){$(c)$}
\end{overpic} 
\begin{overpic}[width=0.4\textwidth]{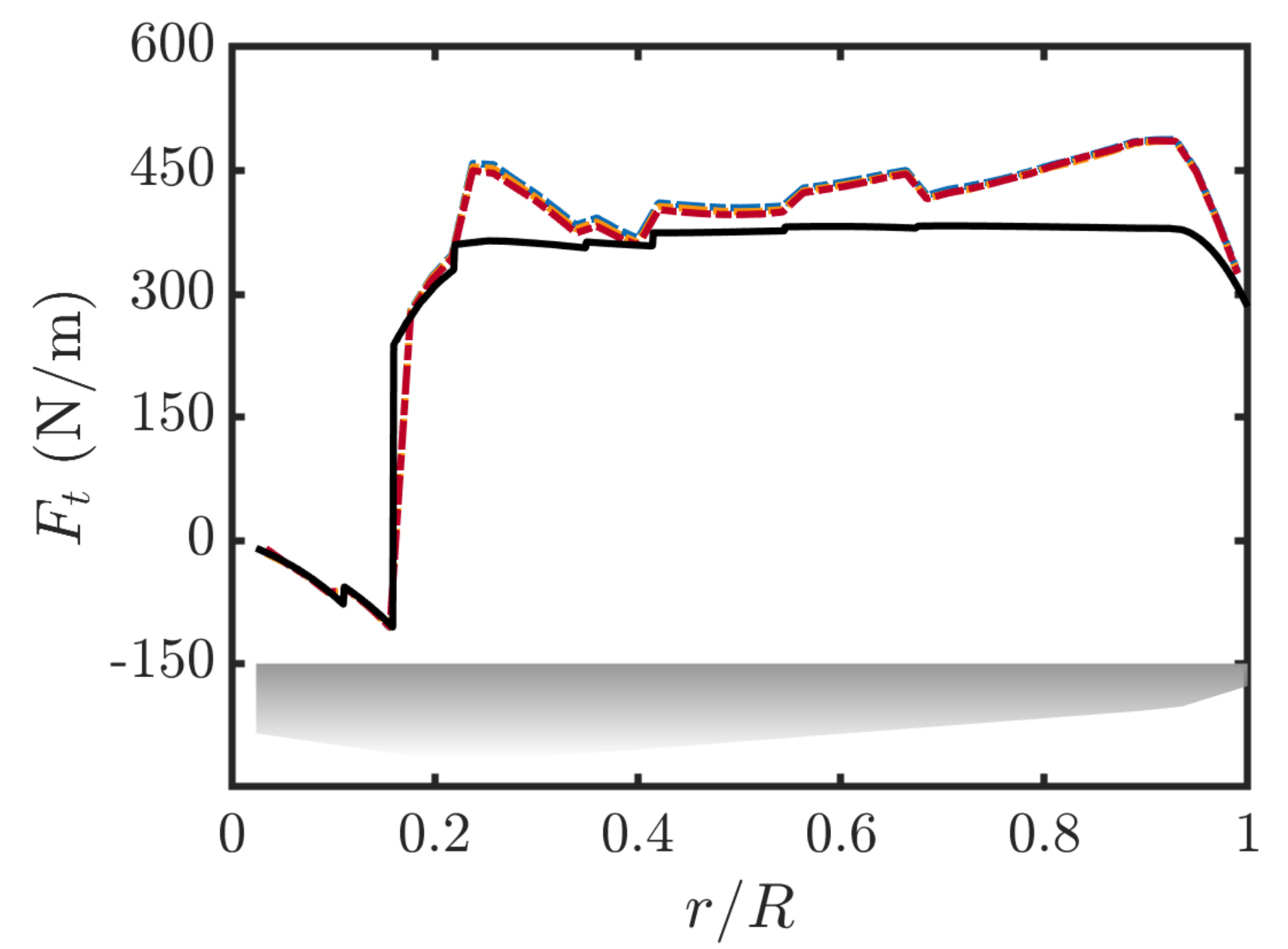}
\put(0,67){$(d)$}
\end{overpic}
\begin{overpic}[width=0.4\textwidth]{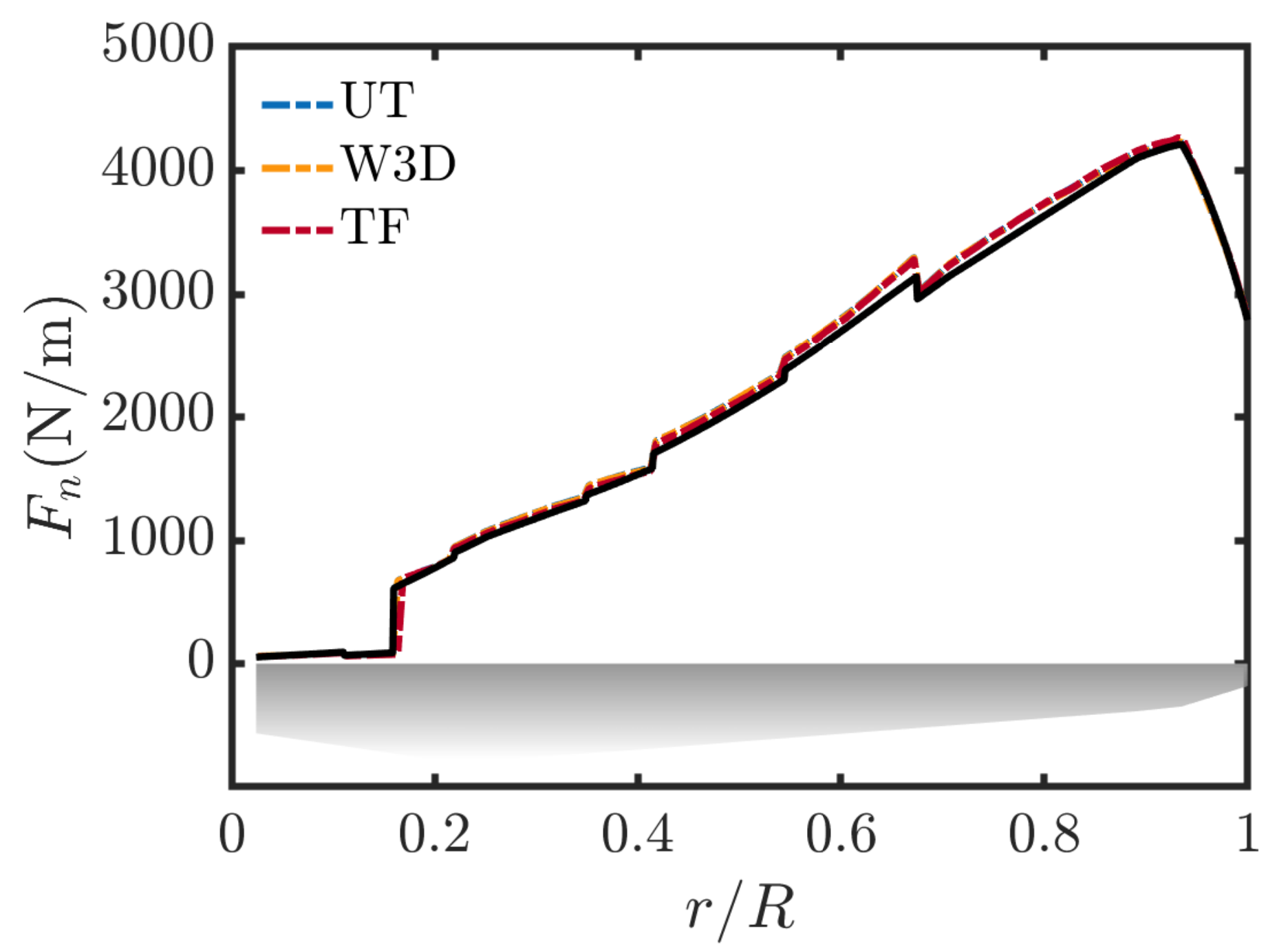}
\put(0,67){$(e)$}
\end{overpic} 
\begin{overpic}[width=0.4\textwidth]{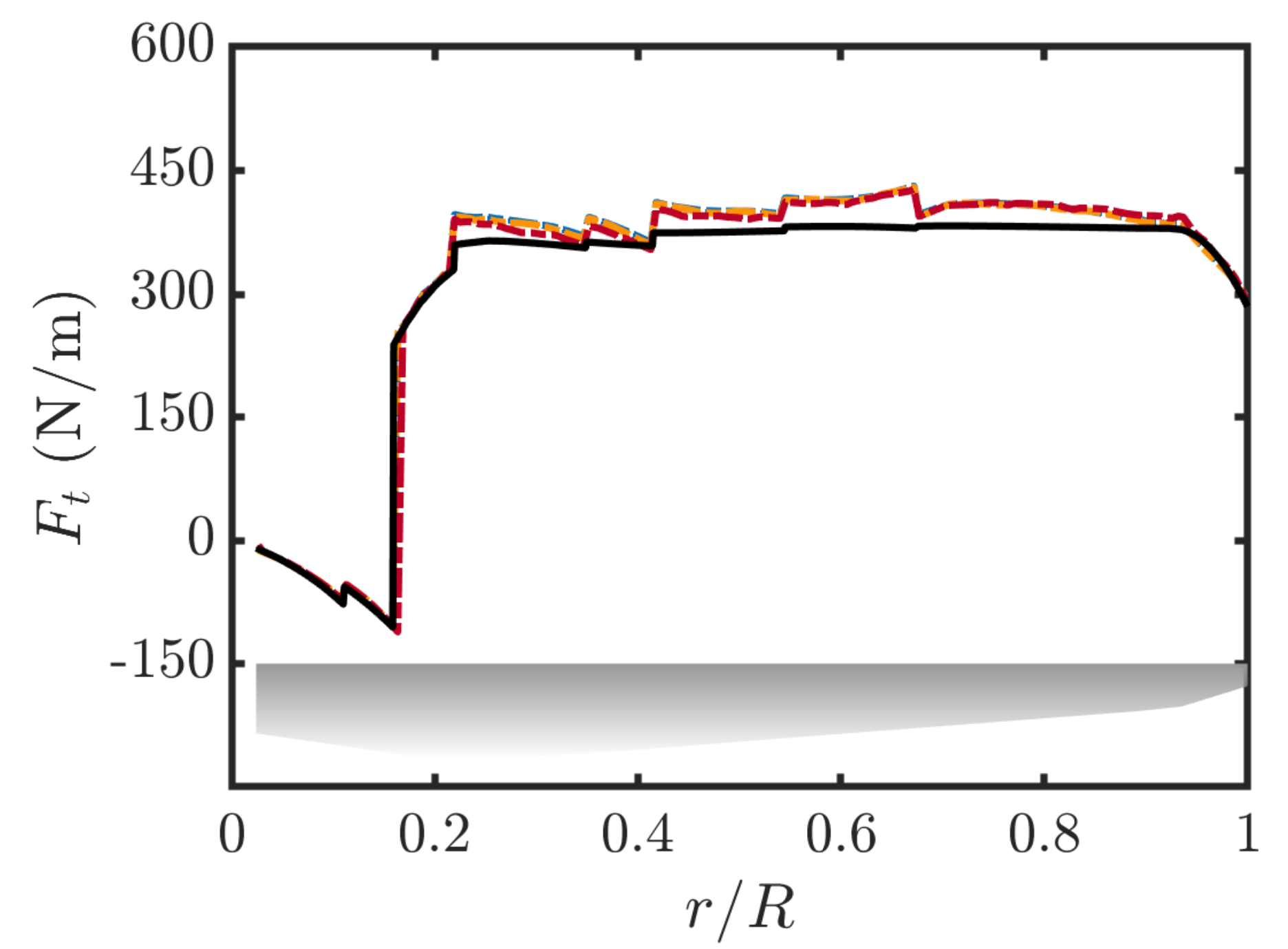}
\put(0,67){$(f)$}
\end{overpic}
\vspace{-12pt} \caption{The (a,c,e) axial and (b,d,f) tangential force per unit span along the blade obtained in the pseudo-spectral LES solver (UT), Winc3D (W3D), and turbinesFoam (TF) simulations for case A (a,b), B (c,d), and F (e,f). The solid line indicates the high-resolution ($N=1024$) BEM reference results and the symbols indicate the simulation results obtained on different grid resolutions, see table \ref{tab.grid} for details.}
\label{fig.grid-blade2}
\end{figure}

\begin{figure}[!t]
\centering
\begin{overpic}[width=0.4\textwidth]{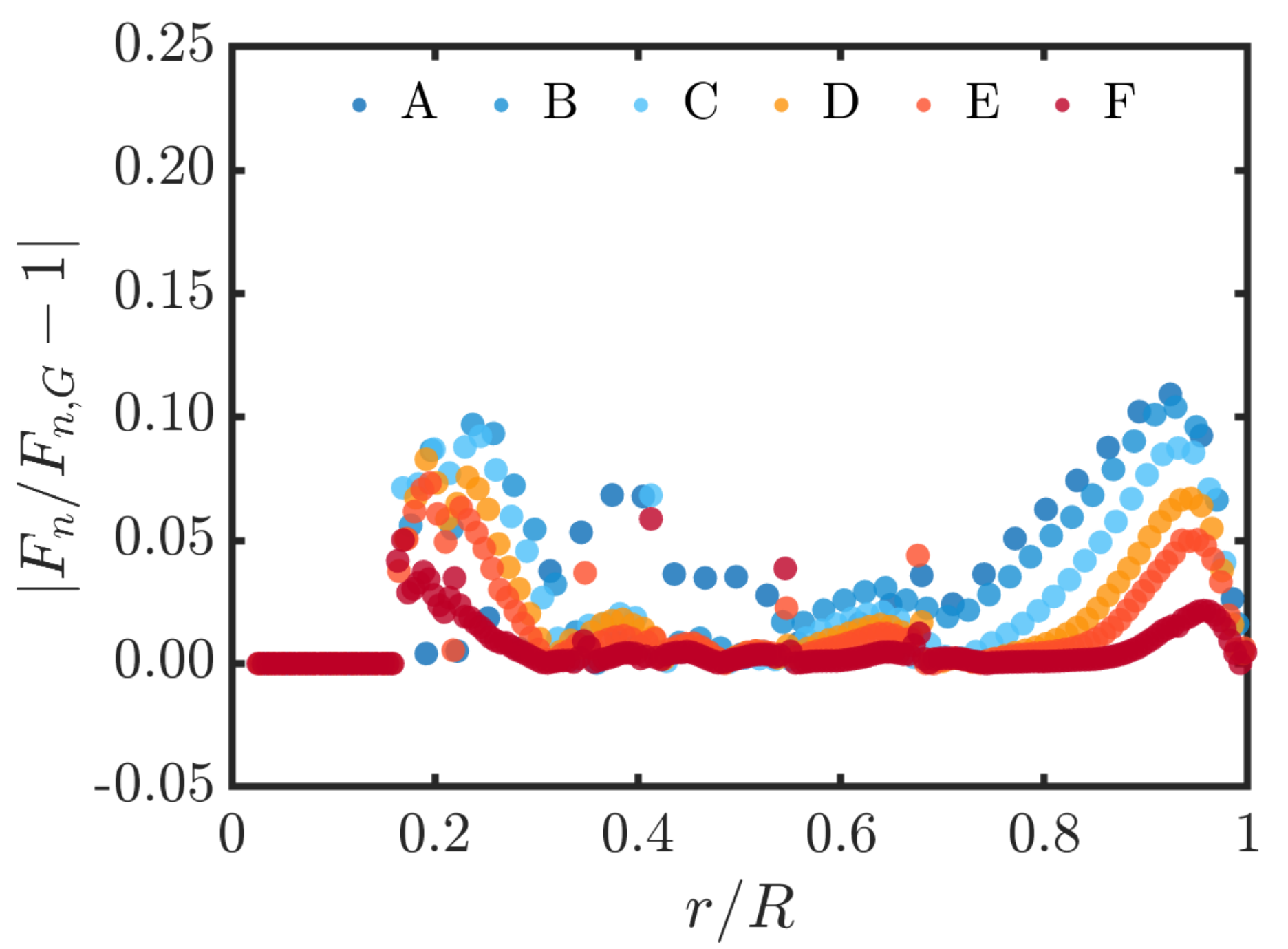}
\put(0,67){$(a)$}
\end{overpic} 
\begin{overpic}[width=0.4\textwidth]{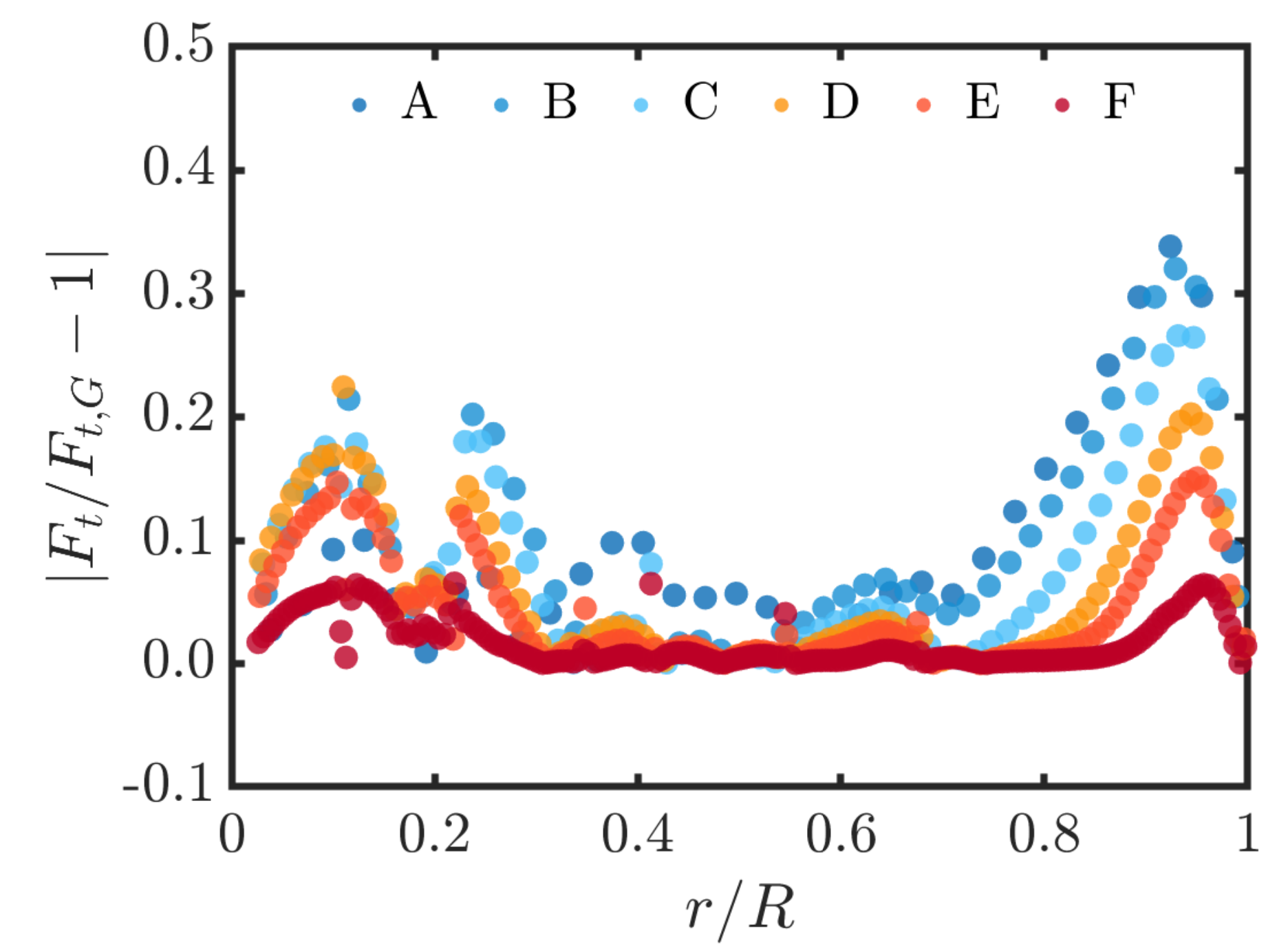}
\put(0,67){$(b)$}
\end{overpic}
\vspace{-12pt} \caption{The relative error of (a) axial and (b) tangential force per unit span along the blade obtained using the pseudo-spectral LES solver. The values of the relative error of axial force at $r/R\le 0.16$ in (a) are set as zero since the axial force therein is negligible small.}
\label{fig.grid-blade-error}
\end{figure}

In this section, we analyze the effect of the grid resolution by varying the number of grid points $N_x \times N_y \times N_z$ in the streamwise, spanwise, and vertical directions, respectively. We ensured that the grid resolution in each direction is the same. The number of actuator points along each blade is $N=N_y/3$. The results are summarized in table~\ref{tab.grid} and figures~\ref{fig.grid-blade1} and \ref{fig.grid-blade2}. The figures compare the axial $F_n$ and tangential $F_t$ per unit span along each blade as a function of the employed grid resolution with high-resolution BEM result obtained using $N=1024$ actuator points. The figure shows that both the axial and tangential forces obtained in the simulations converge towards the BEM results when the resolution of the simulations is increased. Difference between ALM and BEM are observed at the locations where the blade changes abruptly from one airfoil type to another. These discontinuities are most noticeable in the tangential force obtained at lower grid resolutions (Fig.~\ref{fig.grid-blade1}b,d,f). Figure~\ref{fig.grid-blade-error} shows the relative error, defined by the ratio of the results obtained by lower resolution cases (A-F) and the highest resolution case G, of (a) axial and (b) tangential force per unit span along the blade using the pseudo-spectral LES solver. Because the axial force at $r/R\le 0.16$ is negligibly small; the corresponding relative error is defined as zero in figure~\ref{fig.grid-blade-error}(a). The figure reveals that the distribution of the forces along the blades obtained using ALM converge towards the BEM results when the numerical resolution is increased. However, it is essential to note that the local blade forces obtained using an $\sim$2~m resolution and a $\sim$1~m resolution can be as high as $10\%$, see the tangential forces in figure~\ref{fig.grid-blade-error}(a), which indicates that a high resolution is required to capture the local forces on the blades accurately.
We also note that both the axial and tangential forces in most of the middle part of the blade are slightly above the BEM results. This is in agreement with the finding that the total thrust and power output obtained by the ALM simulations are slightly above the BEM theory for the employed grid resolutions, see table \ref{tab.grid}.

\begin{center}
\begin{table}[!t]%
\centering
\caption{Comparison of power $C_P$ and thrust $C_T$ coefficients obtained in the pseudo-spectral LES solver (UT), Winc3D (W3D) and turbinesFoam (TF) simulations. The columns from left to right indicate the case name, the used numerical resolution in the streamwise, spanwise, and vertical direction ($N_x \times N_y \times N_z$), and the grid spacing $\Delta_{\rm grid}$. The number of actuator points along each blade is $N=N_y/3$. The $C_P$ and $C_T$ are obtained using the Gaussian projection function (Eq.~\eqref{eq.gauss}). }
\label{tab.grid}
\begin{tabular*}{500pt}{@{\extracolsep\fill}ccccccccccc@{\extracolsep\fill}}
\toprule
Case & $N_x \times N_y \times N_z$ & $\Delta_{\rm grid}$ (m) & $C_P^{\rm UT}$ & $C_P^{W3D}$ & $C_P^{\rm TF}$ & $C_T^{\rm UT}$ & $C_T^{\rm W3D}$ & $C_T^{\rm TF}$ \\
\midrule
A & $128 \times 96 \times 96$ & 7.88 & 0.6034 & 0.6365 & 0.6355 & 0.8542 & 0.8777 & 0.8764 \\
B & $192 \times 144 \times 144$ & 5.25 & 0.6077 & 0.6032 & 0.6017 & 0.8618 & 0.8584 & 0.8569 \\
C & $256 \times 192 \times 192$ & 3.94 & 0.5891 & 0.5883 & 0.5882 & 0.8521 & 0.8514 & 0.8507 \\
D & $384 \times 288 \times 288$ & 2.63 & 0.5745 & 0.5745 & 0.5717 & 0.8444 & 0.8444 & 0.8415 \\
E & $512 \times 384 \times 384$ & 1.97 & 0.5650 & 0.5655 & 0.5619 & 0.8379 & 0.8382 & 0.8343 \\
F & $768 \times 576 \times 576$ & 1.31 & 0.5547 & 0.5532 & 0.5523 & 0.8313 & 0.8304 & 0.8272 \\
G & $1024 \times 768 \times 768$ & 0.98 & 0.5494 & -- & -- & 0.8280 & -- & -- \\ 
\bottomrule
\end{tabular*}
\end{table}
\end{center}

\begin{figure}[!t]
\centering
\begin{overpic}[width=0.45\textwidth]{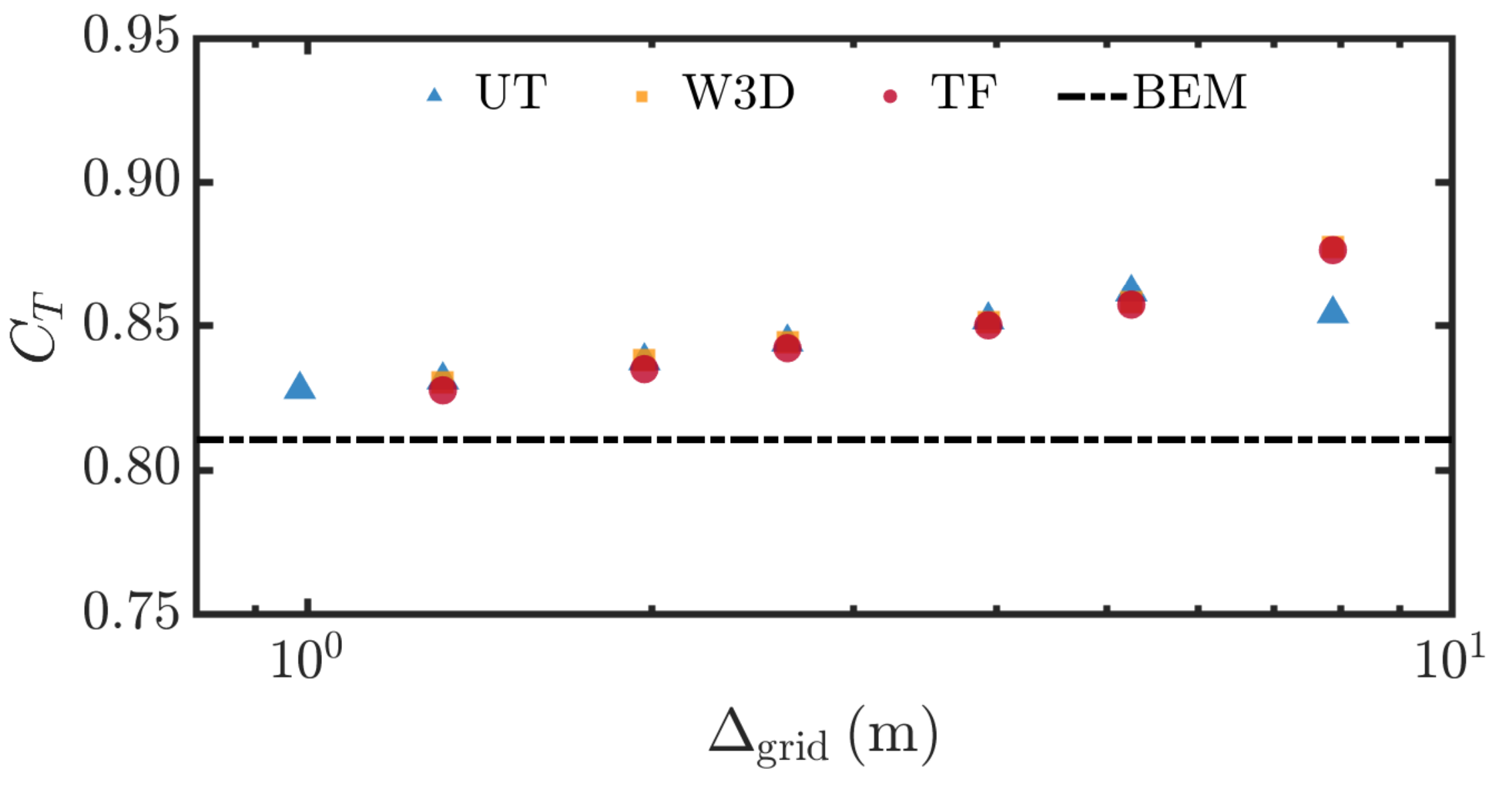}
\put(0,47){$(a)$}
\end{overpic}
\begin{overpic}[width=0.45\textwidth]{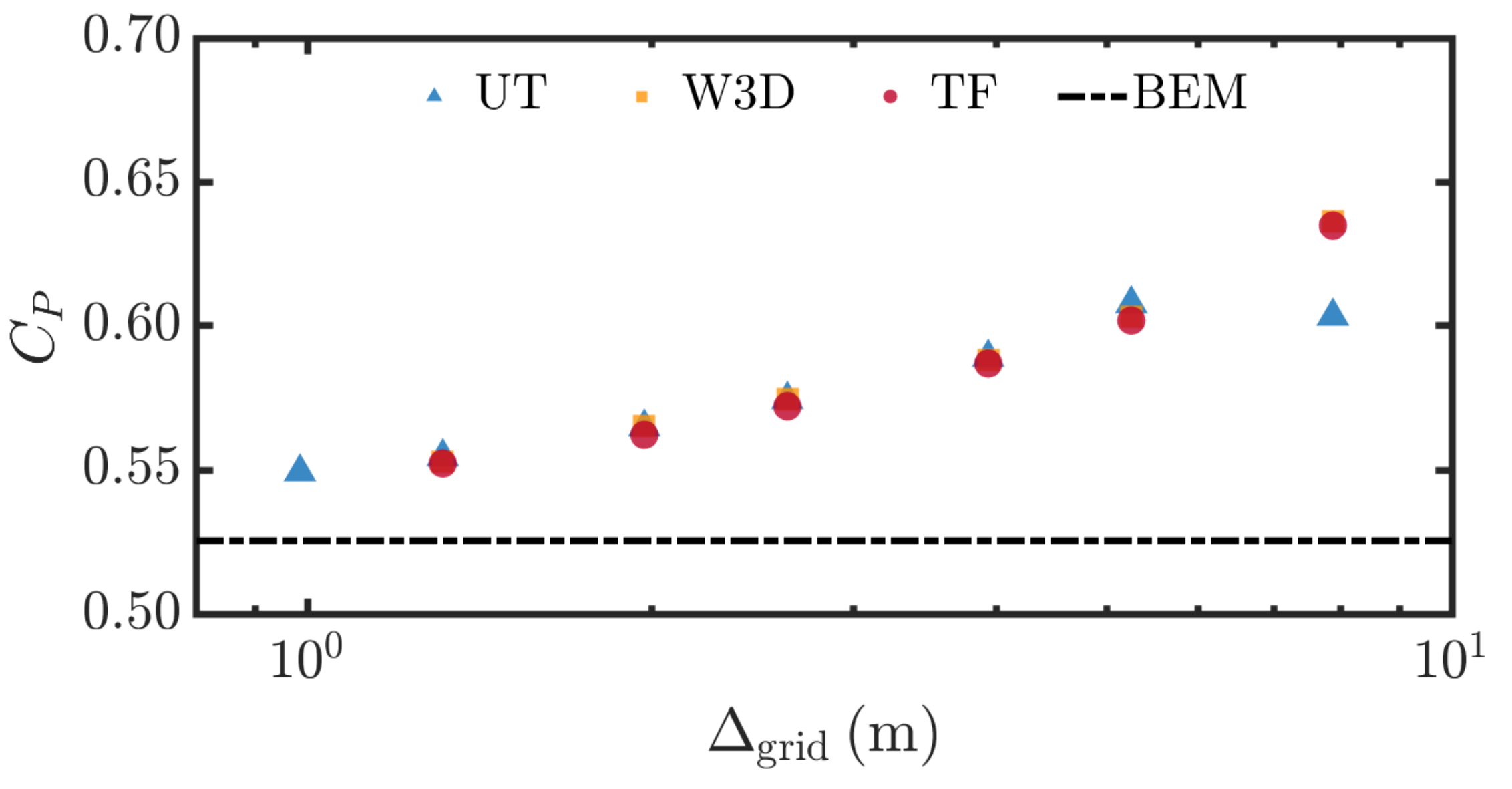}
\put(0,47){$(b)$}
\end{overpic}
\vspace{-12pt} \caption{The (a) thrust $C_T$ and (b) power $C_P$ coefficient of the NREL~5~MW wind turbine as function of the employed grid resolutions $\Delta_{\rm grid}$.
The dashed line indicates the high-resolution ($N=1024$) BEM reference results and the symbols the simulation results, see table \ref{tab.grid} for details.}
\label{fig.grid-power-compact}
\end{figure}

Figure~\ref{fig.grid-power-compact} shows the dependence of the thrust $C_T$ and power $C_P$ coefficient on the employed grid spacing $\Delta_{\rm grid}$. In agreement with the results presented in the table \ref{tab.grid}, the figure shows that the results obtained using the three codes agree very well. In particular, for $\Delta_{\rm grid} \le 5.25$~m the difference is always less than $1\%$. However, for the coarsest mesh $\Delta_{\rm grid} = 7.88$~m a difference of about $6\%$ is observed. We speculate that the differences on this coarsest mesh, which is equivalent to about 8 points per blade length and thus much coarser than what is generally considered acceptable for ALM simulations \cite{lu11, liu21-re}, are due to differences in the employed numerical methods in the three codes. Furthermore, we note the difference in the thrust and power coefficients between the two highest resolution simulations, i.e.\ F ($\Delta_{\rm grid} \le 1.31$~m) and G ($\Delta_{\rm grid} \le 0.98$~m) cases, is less than 1\%, while the difference between the highest resolution case G and the BEM prediction is $4.5\%$ for the power coefficient and $2.1\%$ for the thrust coefficient (see table~\ref{tab.grid}). The power and thrust coefficients in the ALM model are higher than in BEM because the blade forces on coarser meshes are more spread out due to which the wind velocity and forces on the blade are higher at the blade location than on a finer mesh. Therefore, the axial and tangential forces are slightly above the BEM results (see figures~\ref{fig.grid-blade1} and \ref{fig.grid-blade2}).

It is worth mentioning that the obtained LES results are not strictly speaking convergent to the BEM results, even though the difference becomes smaller when the grid resolution increases. For example, the value of $C_T$ obtained by LES with the highest resolution (case G) is about $2.1\%$ larger than the BEM prediction. The physical interpretation of the root cause of these discrepancies lies in the intrinsic difference between BEM and LES. Relevant in the present context are assumptions in BEM theory that each annular ring is independent of every other annular ring and that there is no wake expansion, while these effects are automatically accounted for in LES. On the other hand, ALM in LES uses a force projection method not used in BEM. As $\ep=2.5\Delta_{\rm grid}$ the force projection radius changes with increasing resolution. This means that with increasing grid resolution, the physical problem changes. This is why no absolute convergence of the ALM results is obtained with increasing grid resolution.

\subsection{Effect of the number of actuator points}

\begin{figure}[!t]
\centering
\begin{overpic}[width=0.45\textwidth]{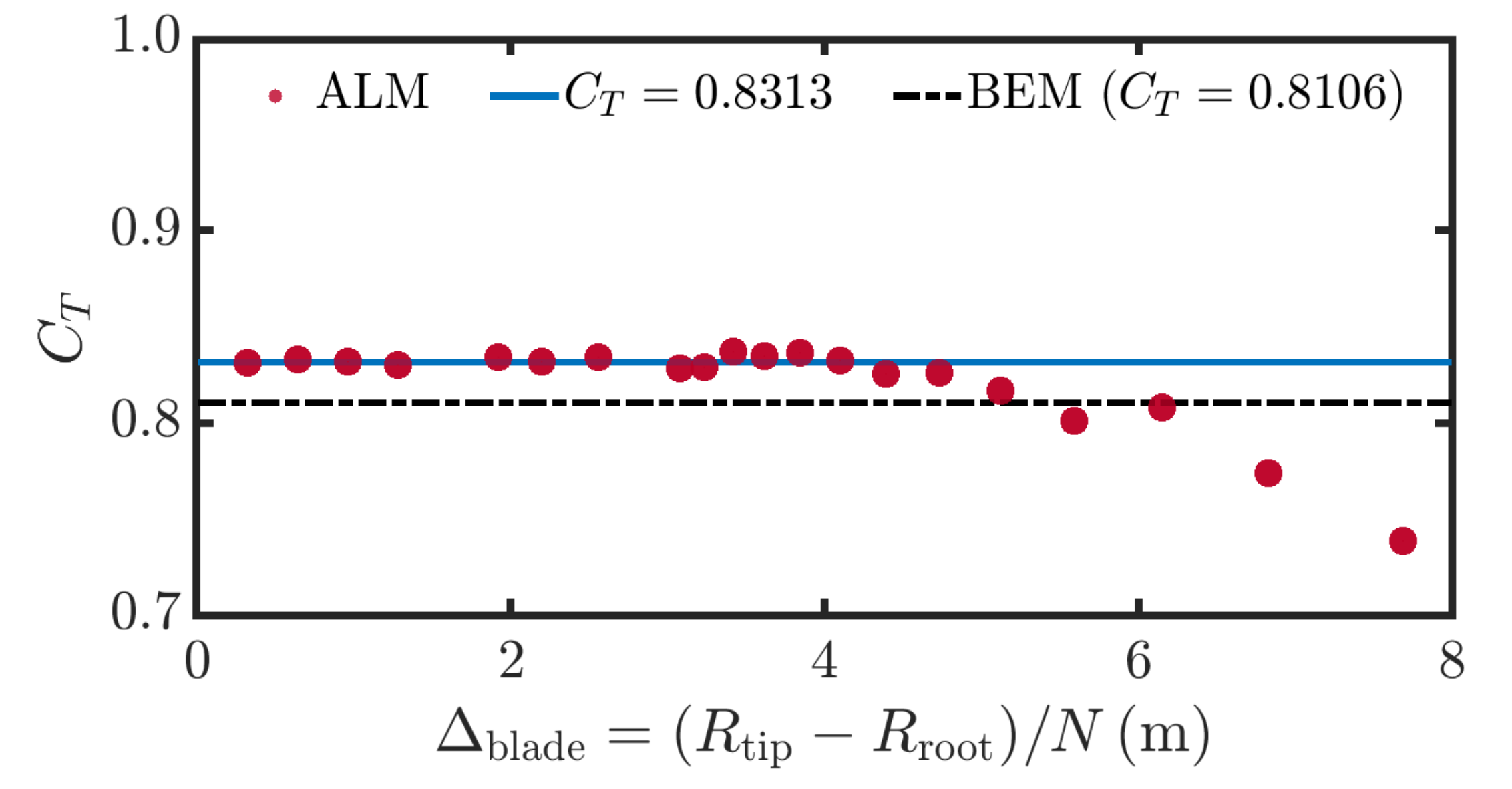}
\put(0,47){$(a)$}
\end{overpic}
\begin{overpic}[width=0.45\textwidth]{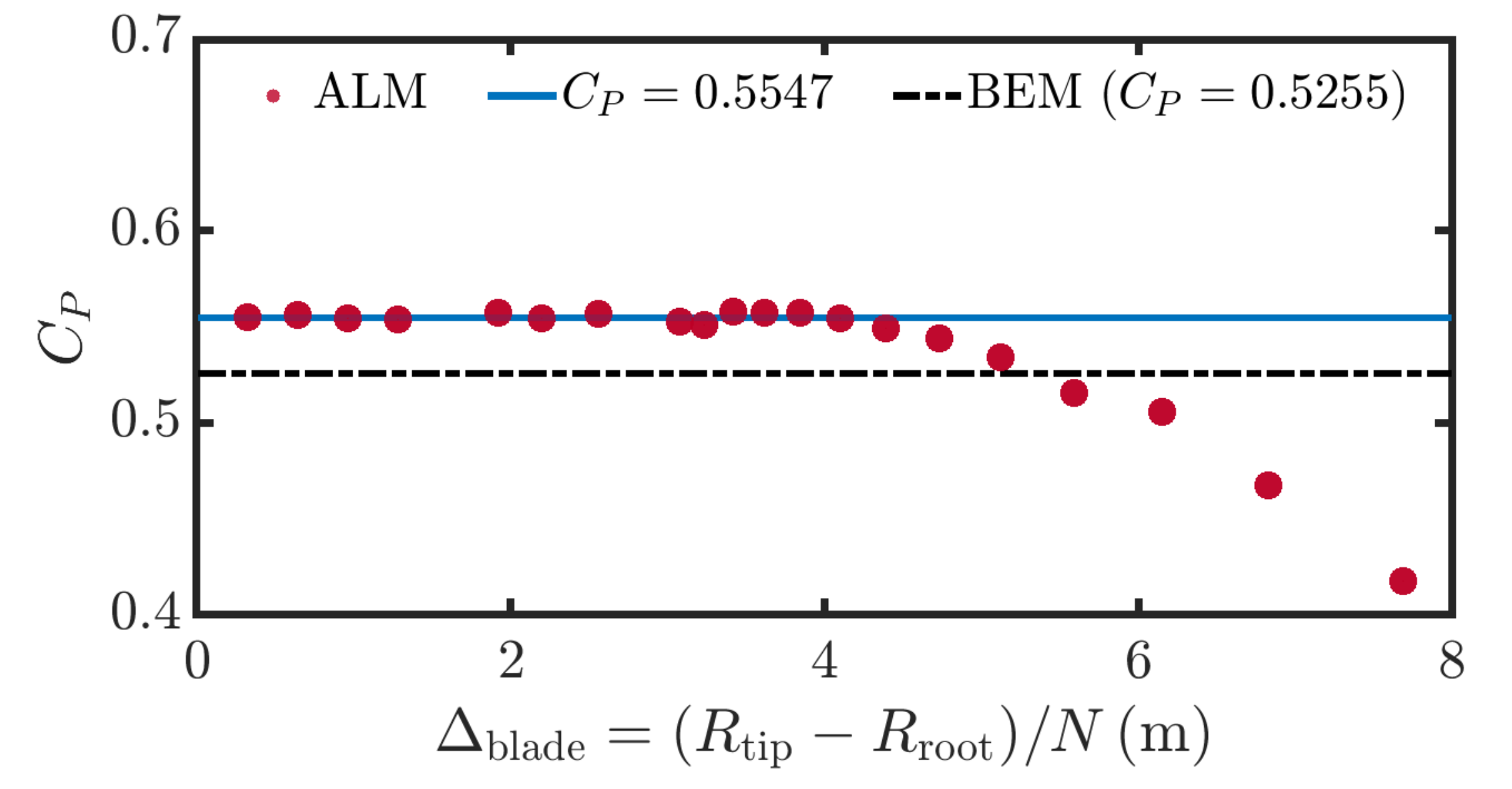}
\put(0,47){$(b)$}
\end{overpic}
\vspace{-12pt} \caption{The (a) thrust $C_T$ and (b) power $C_P$ coefficient as function of the distance between the actuator points $\Delta_{\rm blade}$ for the simulation cases with $\Delta_{\rm grid}=1.31$ meters, see in table \ref{tab.blade}, versus the high-resolution BEM results.}
\label{fig.num-power}
\end{figure}

\begin{figure}[!t]
\centering
\begin{overpic}[width=0.45\textwidth]{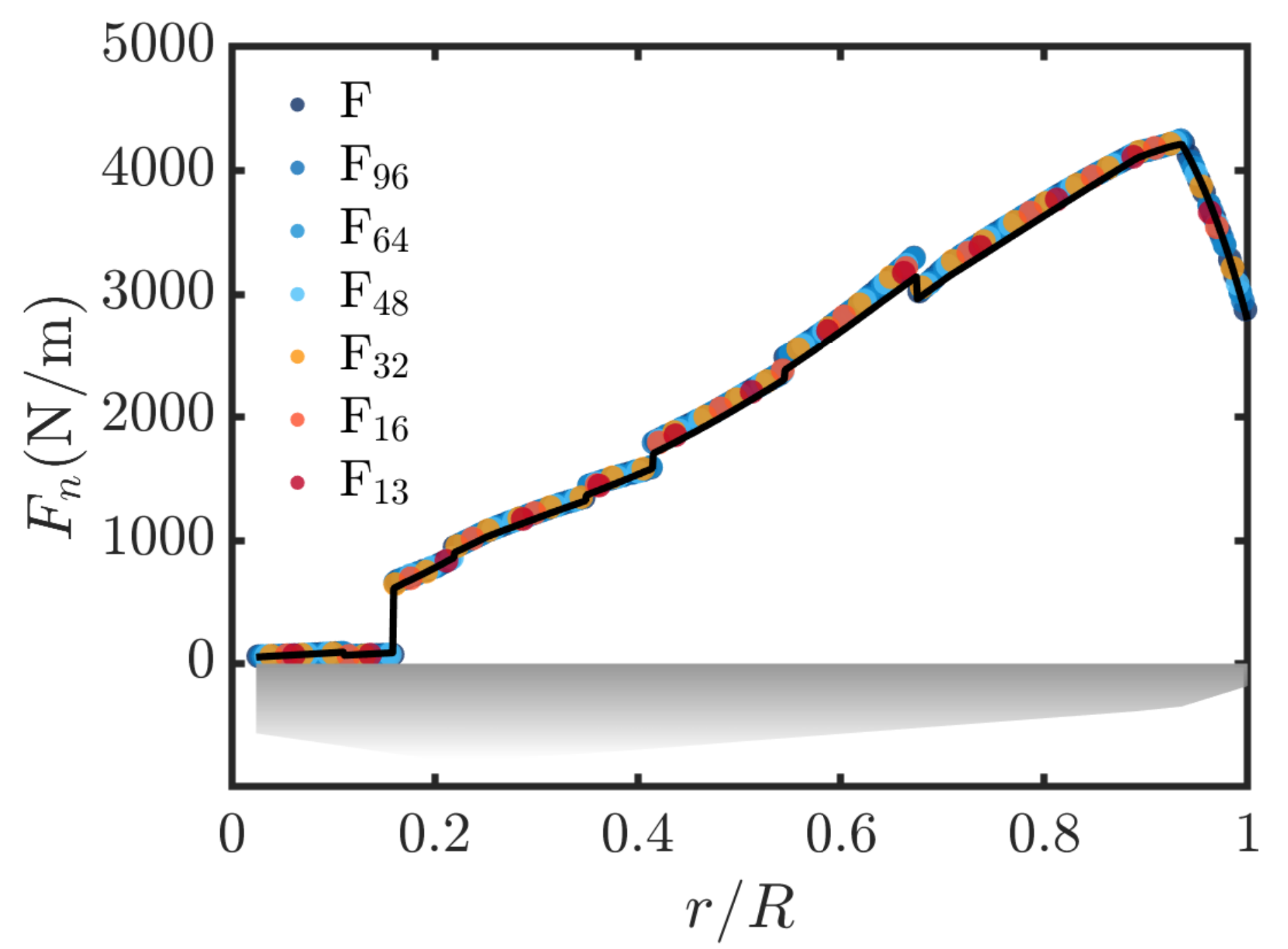}
\put(0,67){$(a)$}
\end{overpic} 
\begin{overpic}[width=0.45\textwidth]{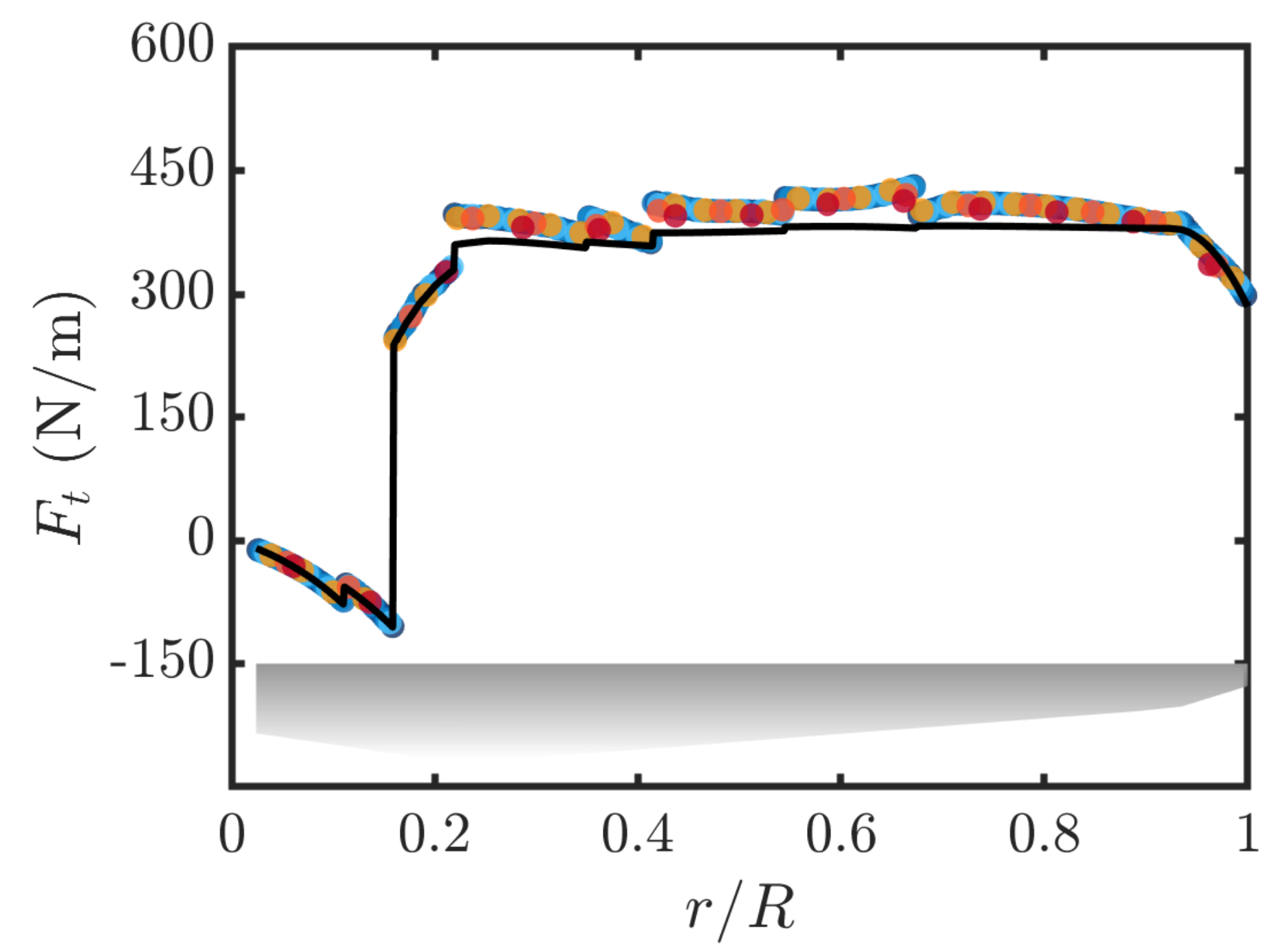}
\put(0,67){$(b)$}
\end{overpic}
\vspace{-12pt} \caption{The (a) axial and (b) tangential force per unit span along the blade. The solid line indicates the high-resolution ($N=1024$) BEM reference results. The symbols indicate the ALM simulation results obtained on grid with $\Delta_{\rm grid}=1.31$ meters but a varying number of actuator points, see table \ref{tab.blade} for details.}
\label{fig.number-blade}
\end{figure}

The BEM analysis in section~\ref{sec.bem} revealed that a limited number of actuator points per blade already allows one to calculate the power and thrust coefficients reasonably accurately. To assess whether this conclusion also holds for a practically implemented ALM model, we systematically vary the number of actuator points as indicated in table~\ref{tab.blade} while keeping the grid resolution constant at $\Delta_{\rm grid} = 1.31$~m. Figure~\ref{fig.num-power} shows that $C_T$ and $C_P$ converge towards the high resolution result obtained on this grid resolution when the distance between the actuator points is $\Delta_{\rm blade} \le 4.1$~m. Figure~\ref{fig.number-blade} confirms that the axial and tangential force per unit span along each blade is captured well, even when the number of actuator points is limited. The convergence around $\sim4$~m corresponds to about three times the grid spacing, which corresponds to the region over which most of the force projection takes place, i.e.\ $d/\epsilon \leq 1.5$ in figure \ref{fig.kernel} on either side of the actuator point. Essentially, this result shows that it is mainly the resolution of the CFD grid $\Delta_{\rm grid}$ that determines the accuracy of the ALM results and that the spacing between the actuator points, or in other words, the number of actuator points per blade, has a limited effect on the accuracy of the power and thrust obtained from the ALM calculations.

\begin{center}
\begin{table}[!t]
\caption{Comparison of the ALM simulations for the NREL~5~MW turbine using the pseudo-spectral LES solver with different number of actuator points $N$ using the Gaussian projection function \er{eq.gauss} and grid resolution $N_x \times N_y \times N_z= 768 \times 576 \times 576$. }
\label{tab.blade}
\begin{tabular*}{500pt}{@{\extracolsep\fill}ccccc@{\extracolsep\fill}}
\toprule
Case & $\Delta_{\rm blade}$ (m) & $N$ & $C_P$ & $C_T$ \\ 
\midrule
F & 0.32 & $192$ & 0.5547 & 0.8313 \\
\midrule
F$_{96}$ & 0.64 & 96 & 0.5558 & 0.8328 \\
F$_{64}$ & 0.94 & 64 & 0.5546 & 0.8316 \\
F$_{48}$ & 1.28 & 48 & 0.5537 & 0.8300 \\
F$_{32}$ & 1.92 & 32 & 0.5573 & 0.8338 \\
F$_{28}$ & 2.20 & 28 & 0.5544 & 0.8320 \\
F$_{24}$ & 2.56 & 24 & 0.5566 & 0.8343 \\
F$_{20}$ & 3.08 & 20 & 0.5528 & 0.8285 \\
F$_{19}$ & 3.24 & 19 & 0.5507 & 0.8287 \\
F$_{18}$ & 3.42 & 18 & 0.5576 & 0.8369 \\
F$_{17}$ & 3.62 & 17 & 0.5575 & 0.8345 \\
F$_{16}$ & 3.84 & 16 & 0.5571 & 0.8365 \\
F$_{15}$ & 4.10 & 15 & 0.5544 & 0.8324 \\
F$_{14}$ & 4.39 & 14 & 0.5494 & 0.8256 \\
F$_{13}$ & 4.73 & 13 & 0.5440 & 0.8259 \\
F$_{12}$ & 5.12 & 12 & 0.5343 & 0.8165 \\
F$_{11}$ & 5.59 & 11 & 0.5157 & 0.8013 \\
F$_{10}$ & 6.15 & 10 & 0.5058 & 0.8079 \\
F$_{09}$ & 6.83 & 9 & 0.4673 & 0.7741 \\
F$_{08}$ & 7.69 & 8 & 0.4179 & 0.7386 \\
\bottomrule
\end{tabular*}
\end{table}
\end{center}

\section{Conclusions}\label{sec.conclusions}

We compared actuator line model (ALM) approach for the NREL~5~MW wind turbine in uniform inflow using three large eddy simulations codes with results from blade element momentum (BEM) theory. The results from the three codes agree within $1\%$ for grid resolution $\Delta_{\rm grid} \le 5.25$~m, which provides cross-validation of the ALM implementations. The ALM results converge towards BEM without the need for tip correction with increasing grid resolution and for $\Delta_{\rm grid} = 0.98$~m the difference between the power and thrust coefficient obtained using ALM and BEM is $4.5\%$ and $2.1\%$, respectively. We note that ALM results are not expected to fully converge towards BEM theory as both methods are slightly different, for example, due to the use of a force projection method in the ALM method. However, we note that the relative difference in the local axial and tangential forces along the blades obtained from ALM simulations using $\Delta_{\rm grid} = 1.97$~m and $\Delta_{\rm grid} = 0.98$~m can be as large as $10\%$, which shows that a high numerical resolution is required to capture local blade loadings accurately. We find that the accuracy of the ALM mainly depends on the employed grid spacing and that reducing the spacing between the actuator points per blade below three times the employed grid spacing has a limited effect on the obtained accuracy due to the force project method employed in the ALM model. The insight that the number of actuator points per blade can be lower than suggested by some previous studies can be helpful to improve the efficiency of simulations in which the ALM overhead is significant. The alternative force projection method that is proposed provides another potential avenue of optimization. This can happen, for example, when simulations are performed on many cores as the ALM calculations are local in space or when the ALM model is employed to simulate the flow in wind farms with various turbines. 

\section*{Acknowledgements}
The work is part of the European Commission Project ``High Performance Computing for Wind Energy (HPCWE)'' with agreement no. 828799. This work is part of the Shell-NWO/FOM-initiative Computational sciences for energy research of Shell and Chemical Sciences, Earth, and Live Sciences, Physical Sciences, FOM, and STW. We also acknowledge the support from FAPESP (Funda{\c c}{\~a}o de Apoio {\`a} Pesquisa do Estado de S{\~a}o Paulo), Proc. 2019/01507-8, for this research. B.~S.~Carmo thanks the Brazilian National Council for Scientific and Technological Development (CNPq) for financial support in the form of a productivity grant, number 312951/2018-3. We acknowledge the following systems, where the simulations in this work were performed: the national e-infrastructure of SURFsara, a subsidiary of SURF corporation, the collaborative ICT organization for Dutch education and research; the Galileo at CINECA, Italy, under a grant from PRACE; the national supercomputer HPE Apollo Hawk at the High-Performance Computing Center Stuttgart (HLRS) under the grant number H2020HPCWE/33766; the SDumont supercomputer, from the National Laboratory for Scientific Computing (LNCC/MCTI, Brazil), through the CADASE project; and the NEXTGenIO system from EPCC -- the University of Edinburgh, also under the HPCWE project.

\end{document}